\documentclass[journal=jctcce,manuscript=art,layout=twocolumn]{achemso}
\usepackage{amsfonts,amsmath,amssymb,mathrsfs}
\usepackage[english]{babel}
\usepackage{booktabs}
\usepackage{cancel}
\usepackage[font=small]{caption}
\usepackage{dsfont}
\usepackage{etoolbox} 
\usepackage{float}
\usepackage[T1]{fontenc}
\usepackage{graphicx}
\usepackage[utf8]{inputenc}
\usepackage[hidelinks]{hyperref}
\usepackage{physics}
\usepackage{setspace}
\usepackage[absolute,overlay]{textpos}
\usepackage{tikz}
\usepackage[normalem]{ulem}
\usepackage{xcolor}
\usepackage{mathtools}
\usepackage{xr}

\usetikzlibrary{arrows.meta,calc,positioning}
\usetikzlibrary{shapes.geometric}

\graphicspath{{Figures/}}

\usepackage[version=3]{mhchem} 

\makeatletter
\patchcmd{\acs@contact@details}{E}{*\,E}{}{}
\makeatother

\newcommand{\deriv}[2]{\frac{\partial #1 }{\partial #2}}

\newcommand{\derivw}[3]{\frac{\partial^2 #1}{\partial #2 \partial #3}}

\newcommand*{\e}{\mathrm{e}}
\newcommand*{\isDefinedAs}{\coloneqq}
\newcommand*{\x}{\boldsymbol{x}}
\newcommand*{\xc}{\boldsymbol{xc}}
\newcommand*{\rv}{\mathbf{r}}
\newcommand*{\xv}{\mathbf{x}}

\newcommand*{\NU}{\boldsymbol{\nu}}
\newcommand*{\ud}{\mathrm{d}}
\newcommand*{\Order}{\mathcal{O}}

\newcommand*{\exa}{\text{exa}}
\newcommand*{\cheap}{\text{cheap}}
\newcommand*{\ext}{\text{ext}}

\newcommand{\figsize}{0.8\textwidth} 
\newlength{\kitzsize}
\let\oldmaketitle\maketitle
\let\maketitle\relax

\author{Nicolas G. Cartier}
\email{n.cartier@vu.nl}
\author{Klaas J. H. Giesbertz}
\affiliation[VU-Amsterdam]{Department of Chemistry \& Pharmaceutical Sciences and Amsterdam Institute of Molecular and Life Sciences (AIMMS), Faculty of Science, Vrije Universiteit, 1081HV Amsterdam, The Netherlands}
\title[SCF-RDMFT-title]{Exploiting the hessian for a better convergence of the SCF RDMFT procedure}
\abbreviations{}

\begin{document}

\twocolumn[
    \begin{@twocolumnfalse}
    \oldmaketitle
    \begin{abstract}
    One-body reduced density matrix functional theory (RDMFT) provides an alternative to Density Functional Theory (DFT), able to treat static correlation while keeping a relatively low computation scaling. Its disadvantageous cost comes mainly from a slow convergence of the self-consistent energy optimisation. To improve on that problem, we propose in this work to use the hessian of the energy, including the coupling term. We show that using the exact hessian is very effective in reducing the number of iterations. However, since the exact hessian is too expensive to use in practice, we propose an approximation based on an inexpensive exact part and BFGS updates.
    \end{abstract}
    \end{@twocolumnfalse}
]


\section{Introduction}
\label{sec:intro}

Density functional theory (DFT) has been widely used to compute the ground state energy of electronic systems over the past few decades because of its advantageous trade-off between cost and accuracy.\cite{GeerlingsDeProft2003,DreizlerGross2012, KohnBecke1996,BearendsGritsenko1997, MardirossianHeadGordon2017,OrioPantazis2009, BartolottiFlurchick1996} However, it can fail already qualitatively, especially for systems with large static correlations.\cite{CohenMoris-Sanchez2008,CohenMoriSanchez2012,BallySastry1997,BraidaPhilippe1998,OssowskiBoyer2003,GruningGritsenko2003,RuzsinskyPerdew2006,WagnerBaker2014} A promising choice to improve the accuracy in these cases, while keeping a reasonable computation scaling is to use one-body reduced density matrix functional theory (RDMFT),\cite{Mazziotti2007,PernalGiesbertz2016, ShadeKamil2017} which naturally describes static correlations.\cite{LathiotakisMarques2008,MentelGritsenko2014, PirisMatxain2010,Piris2014, ShinoharaSharma2015,Piris2017,GibneyBoyn2023} The large cost of the self-consistent field (SCF) minimisation of the ground state energy in RDMFT (SCF-RDMFT) remains, nevertheless, an obstacle to its acceptance by the community.\cite{PirisUgalde2009,PernalGiesbertz2016,LemkeKussmann2022} The large cost is mostly incurred by the exceptionally large number of iterations. The primary focus of this work is then to reduce the number of SCF iterations to bring RDMFT closer as a practical alternative for cases where DFT fails.

DFT relies on the Hohenberg--Kohn theorem,\cite{HohenbergKohn1964} which
proves that the total energy can be written as a functional of the density for $v$-representable densities. A decade later, Gilbert used the same approach to show that the total energy can also be expressed as a functional of the one-body reduced density matrix (1-RDM) $\gamma$, for $v$-representable 1-RDMs.\cite{Gilbert1975} The advantage is that more general perturbations can be considered than only local potentials, e.g.\ magnetic fields, and that the kinetic energy is a simple linear functional of the 1-RDM.
The total energy functional then reads
\begin{multline}\label{eq:RDMFT_E}
    E[\gamma] \isDefinedAs -\frac{1}{2}\int\bigl[\Delta_{\rv}\gamma(\xv,\xv')\bigr]_{\xv' = \xv}\ud \xv \\+ \iint \gamma(\xv,\xv')v_{\ext}(\xv',\xv)\ud\xv\ud\xv' + W[\gamma],
\end{multline}
where $\xv_i = \rv_i\sigma_i$ denotes a combined space-spin coordinate, $\Delta_{\rv}$ the Laplacian with respect to $\rv$, $v_{\ext}$ the (non-)local external potential and $W$ the interaction energy functional, which Gilbert defined as
\begin{align}
W_{\text{G}}[\gamma] \isDefinedAs \langle\Psi[\gamma] \vert \hat{W} \vert \Psi[\gamma]\rangle ,
\end{align}
with $\Psi$ an $N_e$-particle wave function, $\hat{W}$ an electron-electron interaction operator, which is the Coulomb interaction in practice.
Note the similarity with the Hohenberg--Kohn functional in DFT, but also the striking difference that the kinetic energy is not part of the universal functional, so `only' the interaction energy,
for which we do not have an exact explicit expression in terms of the 1-RDM, thus needs to be approximated. 

However, Gilbert's approach suffers from the same problem as the Hohenberg--Kohn theorem: in order to minimise the functional one needs to remain in the domain on which the functional is defined (domain of $v$-representable 1-RDMs\cite{Schilling2018,LiebertChaou2023}), which is unknown in general. This problem was partially solved by Levy,\cite{Levy1979} who extended the functional to the domain of pure state $N$-representable 1-RDMs, i.e.\ 1-RDMs which can be generated by an $N_e$-particle wave function via
\begin{multline}
    \gamma(\xv,\xv') = N_e \int\dotsi\int \Psi(\xv,\xv_2,\dotsc,\xv_N)\\ \cdot\Psi^*(\xv',\xv_2,\dotsc,\xv_N)\dd\xv_2 \dotsb \dd\xv_N.
\end{multline}
The universal functional is then defined via the constrained search construction as
\begin{align}
W_{\text{L}}[\gamma] \isDefinedAs \min_{\Psi \to \gamma}\langle\Psi \vert \hat{W} \vert \Psi\rangle .
\end{align}
Contrary to DFT, a complete characterisation of the domain of pure state $N$-representable 1-RDMs in terms of explicit conditions on the 1-RDM is nonetheless difficult to obtain.\cite{AltunbulakKlyachko2008,Erdahl1987,TheophilouLathiotakis2015,SchillingPittalis2021} Valone therefore generalised Levy constrained search to the domain of ensemble $N$-representable 1-RDMs,\cite{Valone1980}
\begin{align}
W_{\text{V}}[\gamma] \isDefinedAs \min_{\{w_P,\Psi_P\} \to \gamma} \sum_Pw_P\langle\Psi_P \vert \hat{W} \vert \Psi_P\rangle .
\end{align}
The prescription \( \{w_P,\Psi_P\} \to \gamma \) means that the ensemble composed of states $\{\Psi_P\}$ with respective weights $\{w_P\}$ generates the 1-RDM via\cite{Coleman1963}
\begin{multline}
    \gamma(\xv,\xv') = N_e\sum_P \int\dotsi\int w_P\Psi_P(\xv,\xv_2,\dotsc,\xv_N)\\ \cdot\Psi_P^*(\xv',\xv_2,\dotsc,\xv_N)\dd\xv_2 \dotsb \dd\xv_N,
\end{multline}
where $w_P\geq 0$ and $\sum_P w_P=1$.
The functional $W_{\text{V}}[\gamma]$ also has the nice property to be convex.\cite{Lieb1983, PhD-Giesbertz2010}

To ensure that a 1-RDM is ensemble $N$-representable, it is sufficient to impose that $\Tr\{\gamma\}=N_e$, $\gamma$ and $I-\gamma$ are positive semi-definite ($I$, denoting the identity matrix).\cite{Coleman1963}
In order to impose these constraints most SCF procedures in RDMFT work directly with in the basis which diagonalises the 1-RDM\cite{CohenBaerends2002, Pernal2005, PirisUgalde2009, BaldsiefenGross2013}
\begin{equation}
\gamma(\xv,\xv') = \sum_i n_i\phi_i(\xv)\phi^*_i(\xv'),
\end{equation}
where the eigenvalues $n_i$ are called the (natural) occupation numbers (ONs) and the eigenfunctions $\phi_i(\xv)$, the natural orbitals (NOs).\cite{Lowdin1955}
Moreover, most approximate RDMFT functionals are defined in this basis.\cite{GoedeckerUmrigar1998,GritsenkoPernal2005, PirisMatxain2010,PirisLopez2011, Piris2017,Piris2021, Piris2023}

There are several ways to enforce the inequality constraints on the ONs, arising from ensemble $N$-representability, and in particular parametrization by some suitable function of a variable $x$ seems to be a popular choice. One of the earliest parametrizations is the $\cos(x)^2$ function,\cite{Gilbert1975, HerbertHarriman2003, LathiotakisHelbig2007, PirisMitxelena2021} though the Fermi--Dirac function \(1/(\e^{-\beta x} +1)\)\cite{BaldsiefenGross2013, LemkeKussmann2022} and the error function \(\erf(x)\)\cite{YaoFang2021} are perhaps more attractive choices. Other authors opt for more direct optimization with Powell\cite{CohenBaerends2002} or Lagrange multiplier method, in particular also for the constraint on the trace.\cite{LathiotakisMarques2008, PirisUgalde2009}
One can then optimise the ONs using standard methods like conjugate-gradient\cite{LathiotakisHelbig2007,PirisMitxelena2021} or (quasi-)Newton methods.\cite{PirisMitxelena2021,LemkeKussmann2022, YaoFang2021,YaoZhang2022, YaoZhang2022,Lew-YeeFelipe2023}

One also has to impose the orthonormality of the NOs. 
The most popular way, to do so is to optimise the NOs using an iterative diagonalisation of a generalised Fock matrix.\cite{Pernal2005, LathiotakisMarques2008, PirisUgalde2009, BaldsiefenGross2013, LemkeKussmann2022}
Another common way is to write the NOs as a unitary transformation $U$ of an orthonormal basis.
The most standard form is $U=\exp(X)$, taking $X$ as a real antisymmetric matrix.\cite{DouadyEllinger1980,CioslowskiPernal2001,PernalGiesbertz2016}
Some authors also choose to preserve orthonormality only approximately to calculate an optimization step and reorthonormalize the NOs afterwards,\cite{GoedeckerUmrigar2000, CohenBaerends2002, HerbertHarriman2003, PirisMitxelena2021} or brute forcefully apply a Lagrange multiplier method.\cite{Piris2006}

Most approaches described here are based on a two-step scheme in which the ONs and NOs are each optimized in turn. The main disadvantage is that no information of the coupling between the NOs and ONs is used, so typically a large number of macro-iterations (i.e.\ one pass over an ON optimisation followed by an NO optimisation) is needed to obtain consistency between the NOs and ONs. A notable exception investigating a simultaneous optimisation of  NOs and ONs with preconditioned conjugate-gradient method was however published during the review process of the present work,\cite{YaoSu2024} though that work still does not include the coupling between ON-NO within a step.

A second order single step method which optimizes the NOs and ONs simultaneously seems therefore a sensible direction. The coupling between the NOs and ONs is naturally included via the off-diagonal ON-NO blocks of the hessian in a (quasi-)Newton scheme. An additional reason to use the hessian is that the unitary parametrization of the NOs $U(X)$ yields a non-convex energy functional in terms of the parameters $X$ (even if the functional is convex in $\gamma$ e.g.\ the exact RDMFT functional). The energy landscape becomes much more bumpy in $X$ parameter space and the hessian provides very useful information to guide the optimization algorithm through curved valleys.
Since it is not straightforward to include second order information in the iterative diagonalisation approach, we focus in this work to the orthogonalisation of the NOs via a parametrisation by a unitary matrix.

The aim of this work is to reduce the total number of iterations required for the SCF procedure to converge to arbitrary precision. To achieve this goal, we proceed as follows. After giving some details on the implementation, in section~\ref{sec:num_det} and deriving the expression of the exact hessian in section~\ref{sec:exa_hess}, we look at the waste of optimisation steps that can arise when optimising the ONs and NOs separately, in section~\ref{sec:1_step}. Then we try to extract as much information as possible from the exact hessian in section~\ref{sec:cheap_hess}. Third, we introduce an intermediate set of variables, and combine it with our second point to obtain an approximation of the hessian in section \ref{sec:approx_hess}.
We finally conclude in section \ref{sec:conclusion}.

\section{Numerical details}
\label{sec:num_det}

The purpose of this research being to improve the SCF convergence, we want to avoid as much as possible additional difficulties coming from the functionals themselves. To do so, we have decided to test our methods exclusively on the Müller functional, also known as the Buijse--Baerends (BB) functional,\cite{Muller1984,BuijsePhD1991} known for its convexity w.r.t.\ the 1-RDM.\cite{FrankLieb2008} For the Müller functional 
\begin{equation}
    W^{\text{MBB}}[\gamma] = \frac{1}{2}\sum_{ij} n_i n_j [ii|jj] - \frac{1}{2}\sum_{ij} \sqrt{n_i n_j} [ij|ji],
\end{equation}
with
\begin{equation}
[ij|kl] = \int \ud \rv_1 \!\!\int \ud \rv_2 \frac{\phi_{i}(\rv_1)\phi_{j}(\rv_1)\phi_{k}(\rv_2)\phi_{l}(\rv_2)}{\abs{\rv_1-\rv_2}}, 
\end{equation}
where the orbitals $\{\phi_i\}$ can be taken real as we work in a, non-relativistic setting and restricting to the spin-summed case, where $0 \leq n_i \leq 2$, from now on.

As mentioned in the introduction, though the Müller functional is convex w.r.t.\ the 1-RDM, the unitary parametrization $U(X)$ yields a functional which is not convex w.r.t.\ $X$. Thus the hessian is generally indefinite even for this simple functional. It is therefore preferable to turn to methods able to handle indefinite hessians, such as trust-region methods.\cite{HostOlsen2008,HelmichParis2021} We then work with a trust-region (quasi-)Newton algorithm. As the cost of the method will be dominated by the amount of evaluations of the functional and its derivatives, we report all iterations, including the steps rejected by the trust-region algorithm which will be responsible for `jumps' in the energy convergence in the figures. Our implementation is available at \url{https://github.com/NGCartier/SCF-RDMFT_Hess_investigation} and uses PySCF software\cite{SunZhang2020} and the C++ implementation of the \textsc{fides} package,\cite{fides} relying on an algorithm proposed by Coleman and Li.\cite{ColemanLi1994,ColemanLi1996} This method is expensive because of the use of the eigenvalue decomposition of the hessian. However, a trust-region conjugate-gradient method, which is much more affordable, could be used as a more economical alternative.

Throughout this work, we use the Explicit By Implicit (EBI) method\cite{YaoFang2021} to impose the $N$-representability conditions on the ONs. The EBI approach consists in parameterising the ON $n_i =\frac{1}{2}\left(\erf(x_i+\mu)+1\right)\;\forall i$, where $x_i$ are the variables to optimise and $\mu$ is computed to satisfy the constraint on the trace of the 1-RDM. Since many functionals (including the Müller functional) depend on the square root of the ONs, in order to avoid diverging derivatives for $n_i$ going to 0,\cite{CancesPernal2008} we adapt the EBI parametrisation as follows
\begin{equation}\label{eq:EBI}
    \sqrt{n_i} = \frac{1}{\sqrt{2}}\bigl(\erf(x_i+\mu)+1\bigr).
\end{equation}
The NO orthonormality is imposed by a unitary parametrisation via an exponential ($U=\exp(X)$), as explained in the introduction.

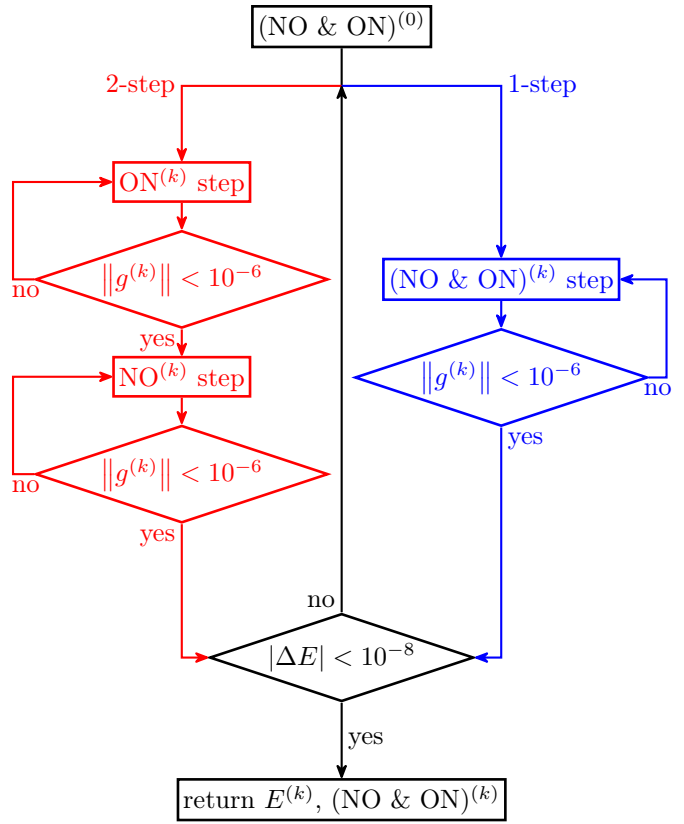
\begin{figure}[t!]
        \begin{tikzpicture}
        [
        every node/.style={node distance=2\kitzsize,line width=0.2\kitzsize, inner sep=2, font=\footnotesize},
        every path/.style={>={Stealth[round]},thick}
        ]
            
            \node[draw](start){(NO \& ON)$^{(0)}$} ;
            \coordinate[below=0.5 of start] (branch_1);
            \draw (start.south) -- (branch_1);
            \begin{scope}[red]
                \node[draw, below left=1 and 1.2 of branch_1](occ_opt){ON$^{(k)}$ step};
                \node[shape aspect=3, draw, diamond, 
                inner sep=1.5](center)[below=of occ_opt]{\(\norm{g^{(k)}} < 10^{-6}\)};
                \node[draw](no_opt) [below=of center]{NO$^{(k)}$ step};
                \node[shape aspect=3, draw, diamond, 
                inner sep=1.5](branch_2)[below=of no_opt]{\(\norm{g^{(k)}} < 10^{-6}\)};
                \draw[->] (occ_opt) -- (center);
                \draw[->] (center) -- node[auto,swap]{yes} (no_opt);
                \draw[->] (no_opt) -- (branch_2);
                \draw[->] (branch_1) -| node[auto,swap]{2-step}(occ_opt);
                \draw[->] (center.west) -| node[auto, pos=0.2]{no}($(occ_opt.west) + (-7.5\kitzsize,0)$) -- (occ_opt);
                \draw[->] (branch_2.west) -| node[auto, pos=0.2]{no}($(no_opt.west) + (-7.5\kitzsize,0)$) -- (no_opt);
            \end{scope}
            \begin{scope}[blue]
                \node[rectangle,draw](1_step) [right=0.7 of center]{(NO \& ON)$^{(k)}$ step};
                \node[shape aspect=3, draw, diamond, 
                inner sep=1.5](branch_1_step)[below=of 1_step]{\(\norm{g^{(k)}} < 10^{-6}\)};
                \draw[->] (1_step) -- (branch_1_step);
                \draw[->] (branch_1_step) -| node[auto,pos=0.25,swap]{no} ($(1_step) + (12.5\kitzsize,0)$) -- (1_step);
                \draw[->] (branch_1) -| node[auto]{1-step}(1_step);
            \end{scope}
            \node[shape aspect=3, draw, diamond, 
            inner sep=1.5](branch_3) [below=7 of branch_1]{\(\abs{\Delta E} < 10^{-8}\)};
            \node[draw](return) [below=1 of branch_3]{return $E^{(k)}$, (NO \& ON)$^{(k)}$};
            \draw[red,->] (branch_2)  |- node[auto, pos=0.05,swap]{yes} (branch_3);
            \draw[blue,->] (branch_1_step) |- node[auto, pos=0.03]{yes} (branch_3.east);
            \draw[->] (branch_3) -- node[auto]{yes} (return);
            \draw[->] (branch_3) -- node[auto, pos=0.02]{no} (branch_1);
        \end{tikzpicture}
	\caption{Algorithm of the 2-step procedure in red, modifications to obtain the 1-step version in blue and common part in black.}
	\label{fig:1step_vs_2step_algo}
\end{figure}

We test our algorithm on alkanes, primary alcohols, the \ce{H2O}, \ce{HF} and \ce{N2} molecules. As example of strongly correlated system we also add a stretched version of the \ce{N2} molecule, at a bond length $R$ equal to height time the equilibrium bond length $R_e$, denoted \ce{N2}($R = 8R_e$). The results are obtained in a cc-pVDZ basis for \ce{H2O}, alkanes and primary alcohols and cc-pVTZ for \ce{HF}, \ce{N2} and \ce{N2}($R = 8R_e$), starting form Hartree--Fock orbitals and a Fermi--Dirac like distribution of the ONs. For the micro-iterations (i.e. single update of the ONs or NOs) we used the gradient,
$\norm{g^{(k)}} < 10^{-6}$ (i.e.\ 2-norm of the gradient), as termination criterion and for the macro-iterations we used the energy difference w.r.t.\ the previous macro-iteration, \(\abs{\Delta E }< 10^{-8}\) for both our implementation of 2-step and 1-step algorithms as depicted in Fig.~\ref{fig:1step_vs_2step_algo}.
To evaluate the convergence of the energy we report at each micro-iteration the energy difference with the best energy $E^{\text{ref}}$ we could obtain for each system.

\section{The exact hessian}
\label{sec:exa_hess}

In the following we use Greek indices for the atomic orbital basis and Latin ones for the NO basis.
Expanding the NOs in a given orbital basis $\{\ket{\chi_{\mu}}\}$ naively as
\begin{align}\label{eq:expOrbAnsatz}
\ket{\phi_i}^{(k+1)} = \sum_{\mu} U_{i\mu}\bigl(Y^{(k+1)}\bigr) \ket{\chi_{\mu}},
\end{align}
(where $Y$ is an anti-symmetric matrix) leads to a very complicated form of the gradient and the hessian, since the derivatives of the unitary matrix $U$ at general $Y$ are very complicated. However, at $Y=0$ the derivatives become quite simple,\cite{CioslowskiPernal2001} so instead, the expansion is only made about the current iterate
\begin{align}\label{eq:curIterExp}
\ket{\phi_i}^{(k+1)} &=\sum_{p\mu}U_{ip}\bigl(X^{(k+1)}\bigr)C^{(k)}_{p\mu}\ket{\chi_{\mu}},
\end{align}
where \(C^{(k)} = U\bigl(X^{(k)}\bigr)C^{(k-1)} \) and \(C^{(0)} = S^{-1/2}\), with $S$ the overlap matrix ($S_{\mu\nu}=\braket{\chi_{\mu}}{\chi_{\nu}}$). Note that \(\exp(X^{(k+1)}) = \exp(Y^{(k+1)})\exp(-Y^{(k)}) \neq \exp(Y^{(k+1)} - Y^{(k)})\) as $Y^{(k+1)}$ and $Y^{(k)}$ do not commute in general.
Expanding the derivatives around the current iterate is exact for an exact energy hessian $H^{\exa}$, but in the case of an approximate hessian more care is needed as it is based on the the difference between gradients at different iterates (see equation~(45) in the supplementary material). In the $\{\ket{\chi_{\mu}}\}$ basis the 1-RDM then attains the form
\begin{equation}\label{eq:1RDM_0th}
    \gamma_{\mu\nu} = \sum_{r s t} C^T_{\mu r}U^T_{rs} n_s U_{st}C_{t\nu}. 
\end{equation}
To derive the gradient and exact hessian, $U(X)$ can be expanded in a Taylor series for common parameterizations (Cayley \cite{Cayley1846}, exponential). For simplicity and consistency with our implementation, we will take the case $U(X) = \exp(X)$, we have $ U_{ij}(X) = 1 + X_{ij} + \frac{1}{2} \sum_{p} X_{ip}X_{pj} + \Order\bigl(X^3\bigr)$. The first and second derivatives are then 
\begin{subequations}
\begin{align}\label{eq:dUdX}
    \left.\deriv{U_{pq}}{X_{ij}}\right\vert_0={}& \delta_{ip}\delta_{jq} - \delta_{iq}\delta_{jp} \\
    \label{eq:ddUddX}
    \left.\derivw{U_{pq}}{X_{ij}}{X_{kl}}\right\vert_0 = {}& \frac{1}{2}\bigl(
         \delta_{ip}\delta_{jk}\delta_{lq} - \delta_{ip}\delta_{jl}\delta_{kq} \notag \\*
    &{}- \delta_{iq}\delta_{jk}\delta_{lp} + \delta_{iq}\delta_{jl}\delta_{kp} \notag \\*
    &{}- \delta_{ik}\delta_{jp}\delta_{lq} + \delta_{ik}\delta_{jq}\delta_{lp} \notag \\*
    &{}+ \delta_{il}\delta_{jp}\delta_{kq} - \delta_{il}\delta_{jq}\delta_{kp} \bigr).
\end{align}
\end{subequations}
Inserting it into the derivative of~\eqref{eq:1RDM_0th} we get
\begin{subequations}
\begin{align}\label{eq:dgammadX}
    \left.\deriv{\gamma_{\mu\nu}}{X_{ij}}\right\vert_0 ={}& (C^T_{\mu j}C_{i\nu} + C^T_{\mu i}C_{ j\nu})(n_i-n_j) \\
    \label{eq:dgammadXX}
        \left.\derivw{\gamma_{\mu\nu}}{X_{ij}}{X_{kl}}\right\vert_0 ={} & 
        \delta_{ik} (C^T_{\mu j}C_{l\nu} + C^T_{\mu l}C_{j \nu}) \notag \\*
        &\cdot(n_i-\frac{1}{2} n_j -\frac{1}{2} n_l) \notag \\*
        &+ \delta_{jl} (C^T_{\mu i}C_{k \nu} + C^T_{\mu k}C_{i \nu}) \notag \\*
        &\cdot(n_j -\frac{1}{2}n_i -\frac{1}{2}n_k) \notag \\*
        &- \delta_{il} (C^T_{\mu j}C_{k \nu} + C^T_{\mu k}C_{j \nu}) \notag \\*
        &\cdot(n_i - \frac{1}{2}n_j - \frac{1}{2}n_k)  \notag \\*
        &- \delta_{jk} (C^T_{\mu i}C_{l \nu} + C^T_{\mu l}C_{i \nu}) \notag \\*
        &\cdot(n_j -\frac{1}{2}n_i -\frac{1}{2}n_l).
\end{align}
\end{subequations}
Many 1-RDM functional approximations are explicitly written in the NO basis because of a dependence $F(n_i,n_j)$ on two ONs.\cite{GoedeckerUmrigar1998,GritsenkoPernal2005,PirisMitxelena2021, PirisLopez2011,Piris2017,Piris2021} The energy functional then takes the form   
\begin{align}\label{eq:E_split}
    E[\gamma] &= \sum_{\mu\nu}\bigg(H^1_{\mu\nu}\gamma_{\nu\mu} + \frac{1}{2}\sum_{\kappa\lambda}\Big(\gamma_{\nu\mu}[\mu\nu|\kappa\lambda]\gamma_{\lambda\kappa} \notag \\ 
    &+ F_{\nu\mu\kappa\lambda}[\mu\lambda|\kappa\nu]\Big)\bigg) \notag \\
    &\eqqcolon E_1[\gamma] + E_H[\gamma] + W_{\xc}[\gamma],
\end{align}
where $H^1= -\frac{1}{2}\Delta + v_{\ext}$ and \(F_{\nu\mu\kappa\lambda} = \sum_{ij} C_{\nu i}^T C_{i\mu}F(n_i,n_j)C_{\kappa j}^TC_{j \lambda}\), the approximated part of the functional. 

From there, we can derive explicit expressions of the entries of $H^{\exa}$ (see Appendix~\ref{sec:appendix_Hexa}).

In the following we will show that the exact hessian provides a good convergence. 
Unfortunately, the excessive cost to compute $H^{\exa}$ of $\Order(N^5)$ (with $N$, number of orbitals) makes its use for practical calculations not a viable option. Therefore, we will aim for an efficient approximation that retains as much of $H^{\exa}$ as possible in sections~\ref{sec:cheap_hess} and~\ref{sec:approx_hess}.

\section{Consistency between NOs and ONs}
\label{sec:1_step}

\begin{figure*}[t]
        \includegraphics[width=\figsize]{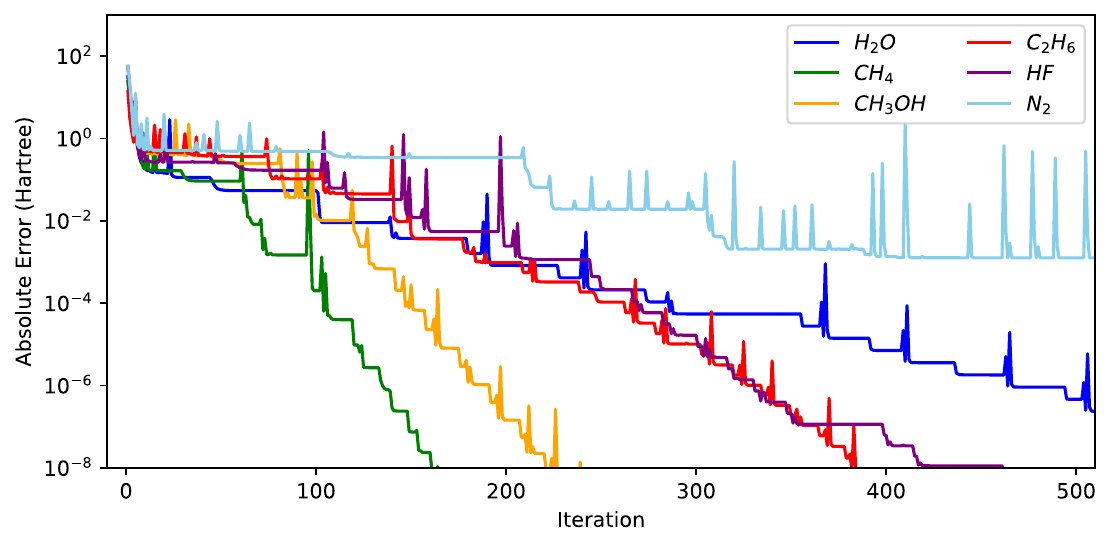}
	\caption{Convergence of the energy ($E^{(k)}-E^{(\text{ref})}$ with respect to the sum of NO and ON iteration) using the 2-step procedure (see text) for different molecules. }
	\label{fig:2step_conv}
\end{figure*}

It is common in RDMFT to optimise the NOs and ONs successively in two separate steps. However, if not done carefully, a 2-step procedure may lead to a waste of computational time, when the implementation keeps optimising the NOs to a high precision, though the change of the ONs in the next step will make such a high precision irrelevant (the reciprocal is also true but the impaired computational cost is negligible). As a demonstration we show the energy convergence at each consecutive micro-iteration step of a 2-step optimization in Fig.~\ref{fig:2step_conv}. The wasteful NO optimization steps is reflected by the plateaus in the energy convergence. It is possible to limit this unwanted behavior by choosing, for example, a gradually tighter termination criterion for the optimisation of the ONs and NOs, but the choice of such procedure can be rather difficult (see supplementary material Sec.~\ref{sec:2step-termination} for details). 

\begin{figure*}[t]
        \includegraphics[width=\figsize]{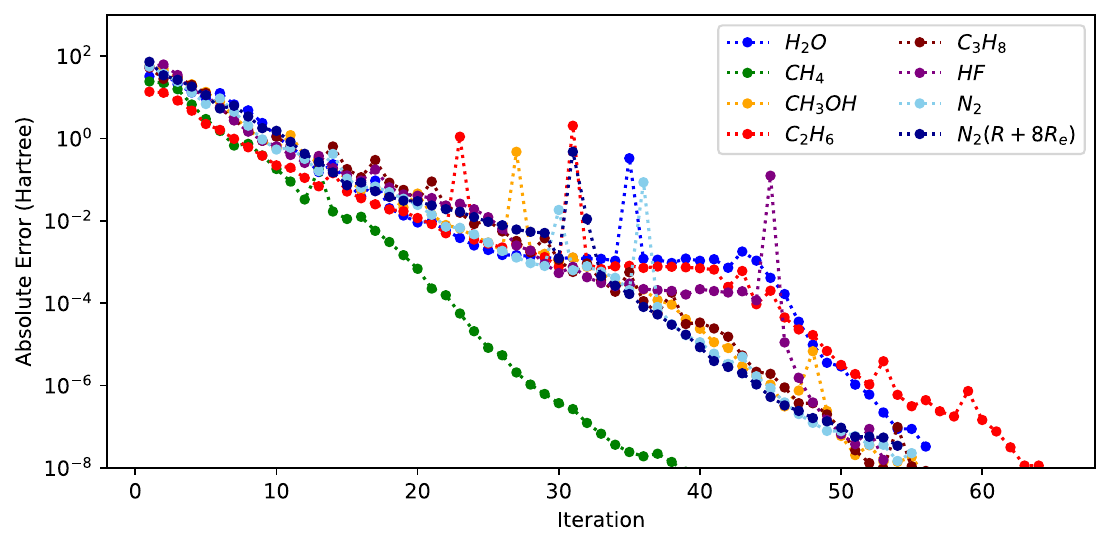}
	\caption{Convergence of the energy ($E^{(k)}-E^{(\text{ref})}$ with respect to the sum of NO and ON iteration) using $H^{\exa}$ for different molecules.}
	\label{fig:exa_hess_conv}
\end{figure*}

The problem of wasteful NO optimization steps is readily circumvented by optimizing both the ONs and NOs simultaneously. An additional advantage is that the coupling between the ONs and NOs can readily be taken into account by including their cross derivative in the hessian.
The results for this 1-step procedure based on the full exact hessian are shown in Fig.~\ref{fig:exa_hess_conv}. The advantage of using the 1-step method is strikingly clear by comparing these results to the 2-step results from Fig.~\ref{fig:2step_conv}.
Indeed, we obtain a significant reduction of the total number of iterations, the energy converging to an error of $5\cdot 10^{-8}$ Hartree for all molecules of the set within 70 iterations. The 2-step procedure, on the other hand, needs a few hundred iterations for the tested molecules. Moreover, the only additional cost per iteration for the 1-step procedure comes  from the coupling between NOs and ONs, which is of order $\Order(N^3)$ only (we have $N$ variables for the ONs). In contrast, the number of entries in the hessian dedicated to NOs is of order $\Order(N^4)$, since the matrix parametrising the NOs, $X$, counts $N(N-1)/2$ degrees of freedom, for a basis of size $N$, which means that the cost of the coupling block is asymptotically negligible compared to the hessian of the NOs.

For a fair comparison, we have tried to keep the implementations as similar as possible, so we have retained the macro-condition of the 2-step algorithm in the 1-step one. In principle, this condition should not play any role in the 1-step case as, it acts as a double check by restarting from the converged point and does indeed not appear when using the exact hessian. We, nonetheless have observed that this restart of the 1-step optimisation can regularly save the convergence as it resets the approximation of the hessian, which may have significantly diverged from the exact one. 

\begin{figure*}[t!]
	\includegraphics[width=\figsize]{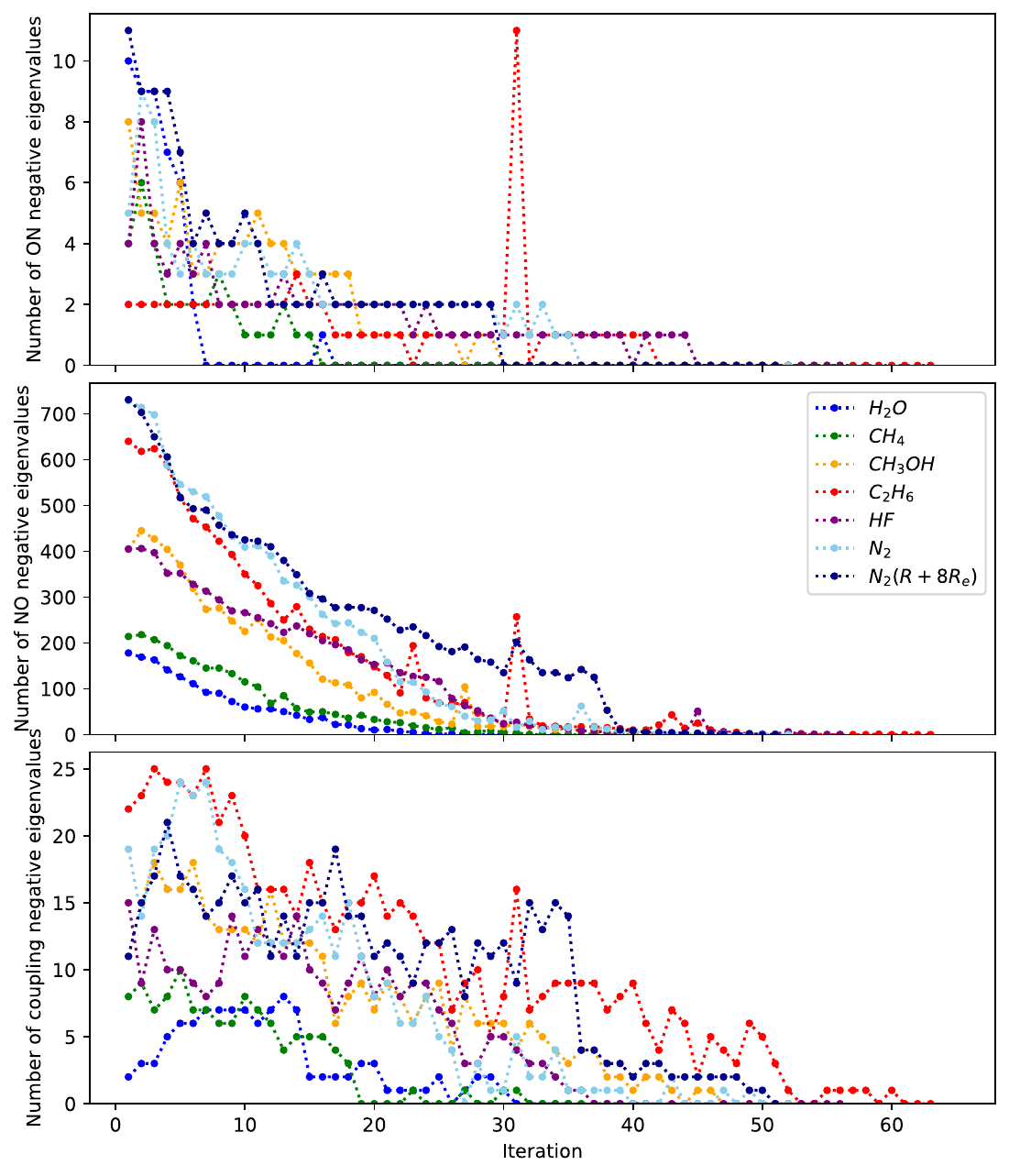}
	\caption{Number of negative eigenvalues of $H^{\exa}$, from the ON block (top panel) the NO block (middle panel) and the total number of negative eigenvalues minus the number of negative eigenvalues from the NO and ON blocks (bottom panel), for different molecules. }
	\label{fig:exa_hess_neigvals}
\end{figure*}

To verify that the 1-step algorithm converged to a minimum, we have computed the eigenvalues of $H^{\exa}$ and checked that they are all non-negative at the end of the convergence. We plot in Fig.~\ref{fig:exa_hess_neigvals} the number of negative eigenvalues of the ON and NO block of the hessian, as well as the difference between the number of negative eigenvalues from the total hessian and and the two aforementioned blocks, which we can interpret as coming from the ON-NO coupling block. Indeed, the number of negative eigenvalues goes down to zero for all tested molecules. Note that the hessian can initially contain a large number of negative eigenvalues, putting the emphasis on the nonconvexity of our problem even for the Müller functional. Moreover, most of the negative eigenvalues come from the NO block of the hessian (middle panel), indicating that the hardest part to handle numerically is the optimisation of the NOs.

\begin{figure*}[t]
        \includegraphics[width=\figsize]{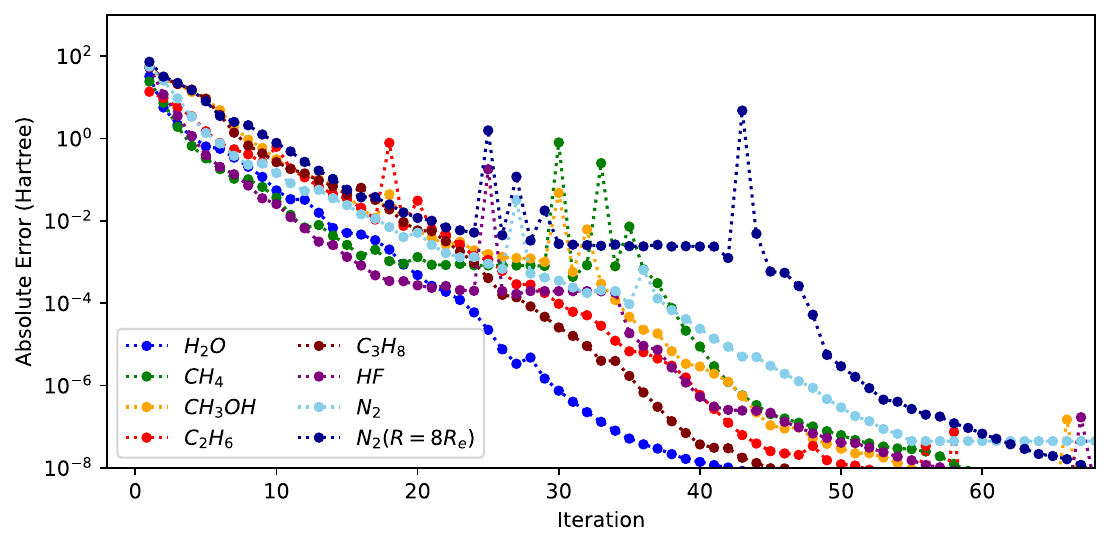}
	\caption{Convergence of the energy ($E^{(k)}-E^{(\text{ref})}$ with respect to micro-iteration count) using the exact ON-ON and NO-NO blocks of the hessian and setting the ON-NO block to 0 for different molecules. }
	\label{fig:exa_hess_no_coupling}
\end{figure*}

One may wonder how important the coupling block of the hessian is, which we included in the 1-step procedure, especially as the number of negative eigenvalues (Fig.~\ref{fig:exa_hess_neigvals}) indicates that the most difficult part is actually the NO block. To investigate this point, we have tested the 1-step algorithm with the coupling block of the hessian set to zero. The results are shown in Fig.~\ref{fig:exa_hess_no_coupling}.

Comparing Figs.~\ref{fig:exa_hess_conv} and~\ref{fig:exa_hess_no_coupling}, we observe that the convergence is roughly similar for most molecules, except for \ce{N2}. Somewhat surprisingly, \ce{N2} at equilibrium distance is more challenging\footnote{The molecule \ce{N2} only converged to~$3\cdot 10^{-8}$ Hartree in the first macro-iterations which had 153 micro cycles and after stopping at the second macro iteration it only converged to $1.2\cdot 10^{-8}$ Hartree.} than stretched \ce{N2} which is typically considered to be more challenging due to its strong correlation character. The main difference of \ce{N2} w.r.t.\ the other systems is that some of the significantly fractionally occupied NOs are (nearly-)degenerate. It seems that the NO-ON coupling block is very helpful to handle those.
However, for some systems (\ce{H2O}, \ce{C2H6} and \ce{C3H8}) setting the coupling block to zero yields faster convergence. Overall it appears that, the coupling block is important to make the 1-step method robust, but does not affect the convergence rate very significantly. For an approximate hessian, the coupling block seems to be somewhat more significant (see supplementary material Sec.~\ref{sec:couplingApproxHes}).

\section{Extracting the cheap part of the exact hessian}
\label{sec:cheap_hess}

\begin{figure*}[t]
        \includegraphics[width=\figsize]{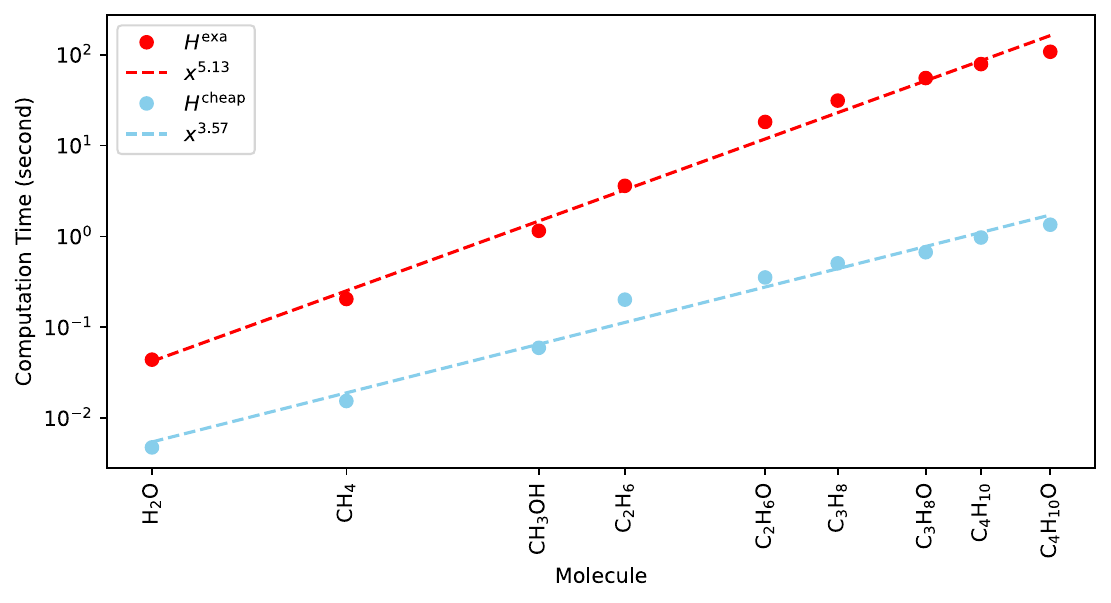}
	\caption{Computation time to compute the exact hessian $H^{\exa}$ (red dotes) and cheap part of the hessian $H^{\cheap}$ (blue dotes) w.r.t.\ the size of different molecules. Asymptotic behaviours are fitted (dashed lines), the formal scaling is $\Order(N^5)$ for $H^{\exa}$ and $\Order(N^4)$ for $H^{\cheap}$.}
	\label{fig:hess_cost}
\end{figure*}

For functionals of the form $F(n_i,n_j) = f(n_i)f(n_j)$, sometimes called separable functionals, we can use the idea proposed by Giesbertz\cite{Giesbertz2016} to reduce the scaling of the contribution to $H^{\exa}$ from the first term of~\eqref{eq:ddExcddx}, first two terms of~\eqref{eq:ddExcdXdx} and the first height lines of~\eqref{eq:ddExcddX}, from $\Order(N^5)$ to $\Order(N^4)$ (see Fig.~\ref{fig:hess_cost}). Indeed, denoting $v^K_{ij} = \sum_{p} f(n_p) [ip|jp]$, \eqref{eq:ddExcddx}, \eqref{eq:ddExcdXdx} and~\eqref{eq:ddExcddX} become
\begin{equation}\label{eq:ddExcddx_sep}
    \left.\derivw{W_{\xc}}{x_i}{x_j}\right\vert_0 = \sum_k\left(\derivw{f(n_k)}{x_i}{x_j}v^K_{kk} + \deriv{f(n_k)}{x_i} \deriv{v^K_{kk}}{x_j} \right),
\end{equation}
\begin{align}\label{eq:ddExcdXdx_sep}
    \left.\derivw{W_{\xc}}{X_{ij}}{x_k}\right\vert_0 = &2\bigg( \deriv{f_i}{x_k}v^K_{ij} - \deriv{f_j}{x_k}v^K_{ji} \notag\\* 
    &+ f_i \deriv{v^K_{ij}}{x_k} - f_j\deriv{v^K_{ji}}{x_k}\bigg).
\end{align}
and 
\begin{equation}\label{eq:ddExcddX_sep}
    \begin{split}
    \left.\derivw{W_{\xc}}{X_{ij}}{X_{pq}}\right\vert_0 &= 
        2\delta_{pi}v^K_{qj}(2f(n_i)-f(n_j)-f(n_q)) \\
        & + 2\delta_{qj} v^K_{ip}(2f(n_j)-f(n_i)-f(n_p))\\
	& - 2\delta_{iq} v^K_{jp}(2f(n_i)-f(n_j)-f(n_p)) \\
        & - 2\delta_{jp} v^K_{iq}(2f(n_j)-f(n_i)-f(n_q)) \\
	& + 4\big(f(n_i) f(n_p) -f(n_j) f(n_p) \\
        &-f(n_i) f(n_q) + f(n_j) f(n_q)\big)[ip|jq].	
    \end{split}
\end{equation}
We can analogously compute the corresponding terms of~\eqref{eq:ddEHddx}, \eqref{eq:ddEHdXdx} and~\eqref{eq:ddEHddX} in $\Order(N^4)$, by computing separately $v^J_{ij} = \sum_{p} f(n_p) [ij|pp]$. Moreover, the one-electron part of $H^{\exa}$ can clearly be obtained in $\Order(N^4)$ [even only $\Order(N^3)$].

For the following expressions, we use uppercase for indices going from $N+1$ to $N(N+1)/2$, which can more conveniently be mapped to two indices (the first one going from $1$ to $N$ and the second one being strictly lower than the first one).
We denote $H^{\exp}$ such that
\begin{subequations}
\begin{align}\label{eq:exp_xpression_dxdx}
    H^{\exp}_{ij} ={}& \sum_k\left(\deriv{f(n_k)}{x_i} \deriv{v^J_{kk}}{x_j} -\deriv{f(n_k)}{x_i} \deriv{v^K_{kk}}{x_j} \right) \\
\intertext{for the ON block,}
\label{eq:exp_xpression_dxdX}
    H^{\exp}_{Ik} ={}& H^{\exp}_{ijk}= 2\bigg(f(n_i) \deriv{v^J_{ij}}{x_k} - f(n_j)\deriv{v^J_{ji}}{x_k}  \notag\\*
    &- f(n_i) \deriv{v^K_{ij}}{x_k} + f(n_j)\deriv{v^K_{ji}}{x_k}\bigg) \\
\intertext{for the coupling block and}
\label{eq:exp_xpression_dXdX}
        H^{\exp}_{IP} ={}& H^{\exp}_{ijpq}= 4(n_i n_p - n_j n_p - n_i n_q + n_j n_q)[ij|pq] \notag \\*
        &{} + 4\big(f(n_i)f(n_p) -f(n_j)f(n_p) \notag \\*
        &-f(n_i)f(n_q) + f(n_j)f(n_q)\big)[ip|jq]
\end{align}
\end{subequations}
for the NO block and $H^{\cheap} = H^{\exa}- H^{\exp}$. The scaling to obtain $H^{\cheap}$ is then $\Order(N^4)$ for a separable functional (that is, the same as to obtain the gradient), and we want to approximate $H^{\exp}$ only.

\begin{figure*}[t!]
        \includegraphics[width=\figsize]{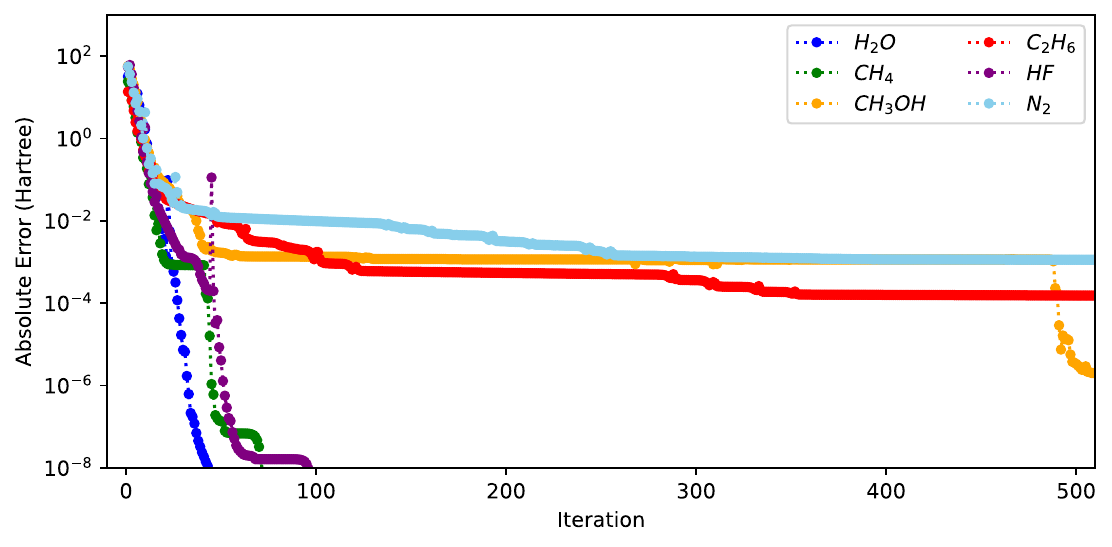}
	\caption{Convergence of the energy ($E^{(k)}-E^{(\text{ref})}$ with respect to micro-iteration count) using $B = H^{\cheap}$ for different molecules. }
	\label{fig:cheap_hess_conv}
\end{figure*}

\begin{figure*}
        \includegraphics[width=\figsize]{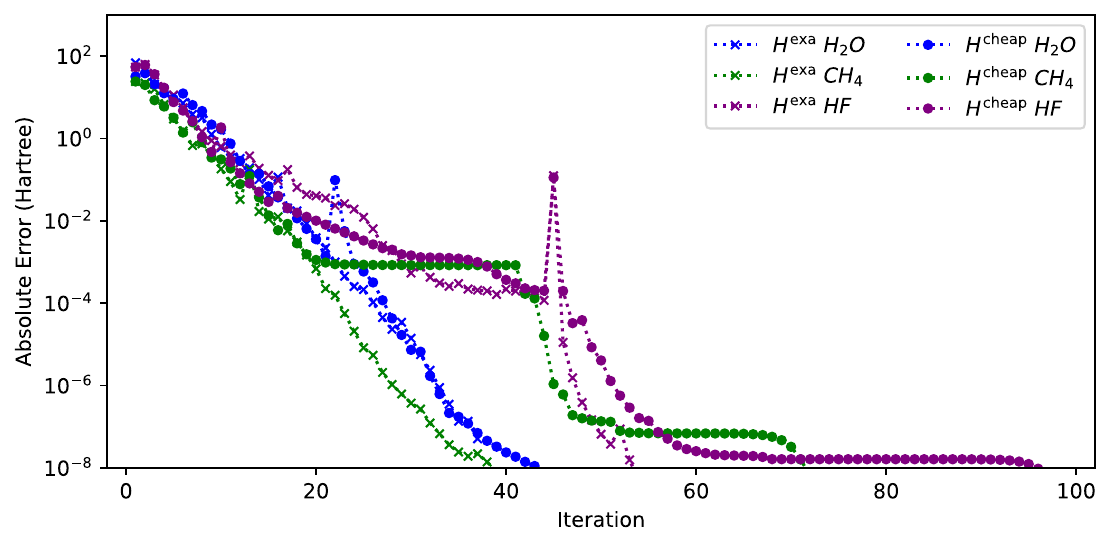}
	\caption{Comparison of the convergence of the energy ($E^{(k)}-E^{(\text{ref})}$ with respect to micro-iteration count) using $B = H^{\cheap}$, denoted by $\bullet$ and $B = H^{\exa}$, denoted by $\times$ for different molecules. }
	\label{fig:cheap_vs_exa_hess_conv}
\end{figure*}

To do so, we can first neglect $H^{\exp}$ entirely and take $B=H^{\cheap}$ as an hessian approximation. (We use $B$ to denote an approximation to the hessian). We tried it for our set of molecules in Fig.~\ref{fig:cheap_hess_conv} and observed a good convergence for the first few tens of iterations. This approximation can even accidentally outperform the exact hessian (see the \ce{H2O} molecule in Fig.~\ref{fig:cheap_vs_exa_hess_conv}). However, this is a very crude approximation and the algorithm cannot converge for half of the molecules. This seems to indicate that $H^{\cheap}$ provides a good approximation of $H^{\exa}$ for the first few iterations which is sufficient to make \ce{H2O}, \ce{CH4} and \ce{HF} converge up to $10^{-8}$ Hartree but only $10^{-2}$ for \ce{N2} and $10^{-4}$ for \ce{C2H6}.

\begin{figure*}[t!]
    \centering
    \begin{minipage}{\textwidth}
        \includegraphics[width=0.5\textwidth]{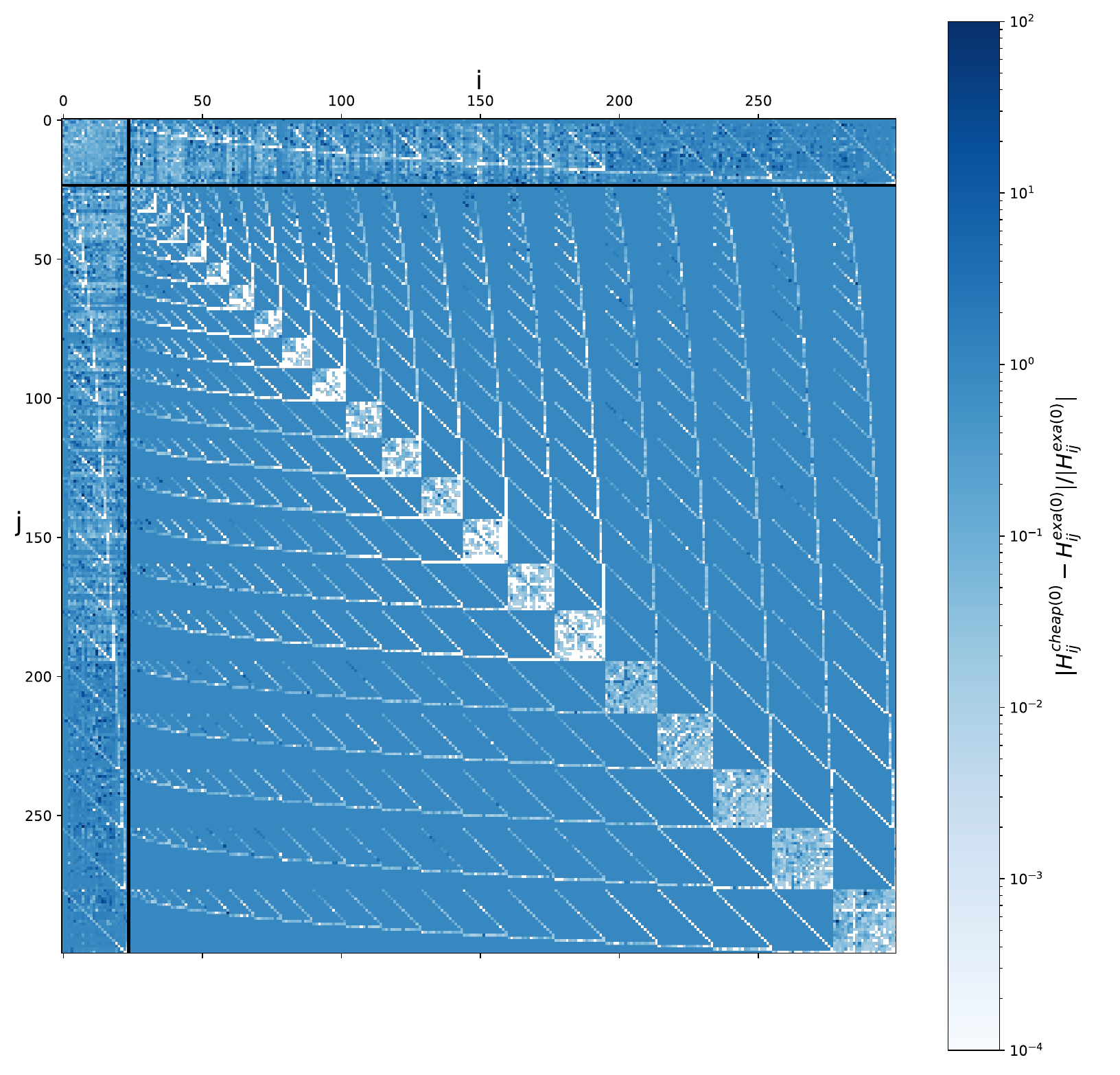}
        \includegraphics[width=0.5\textwidth]{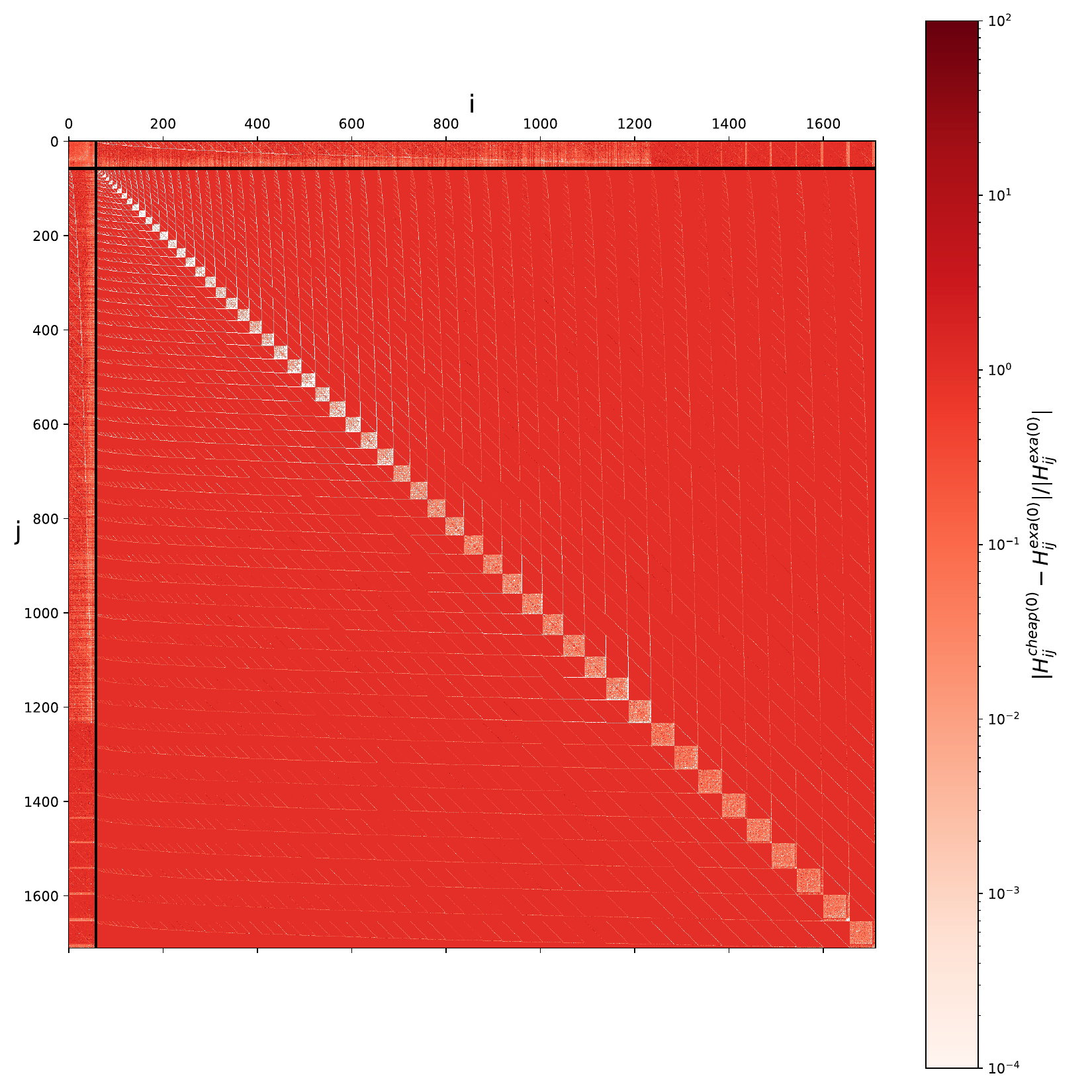}
        \\
        \includegraphics[width=0.5\textwidth]{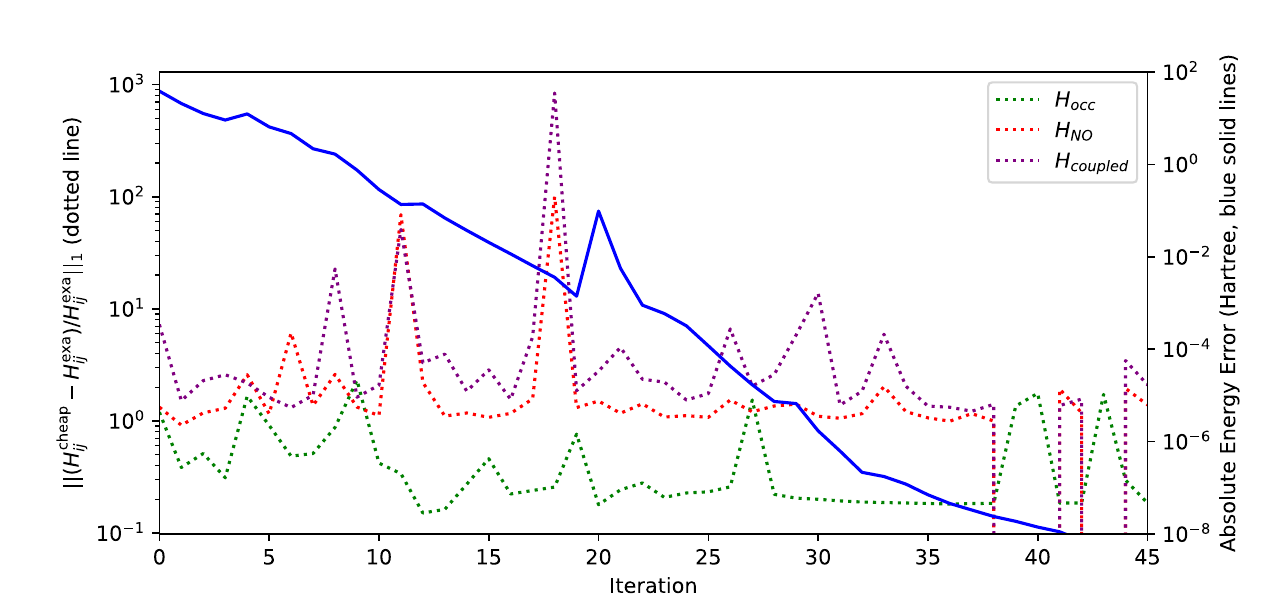}
        \includegraphics[width=0.5\textwidth]{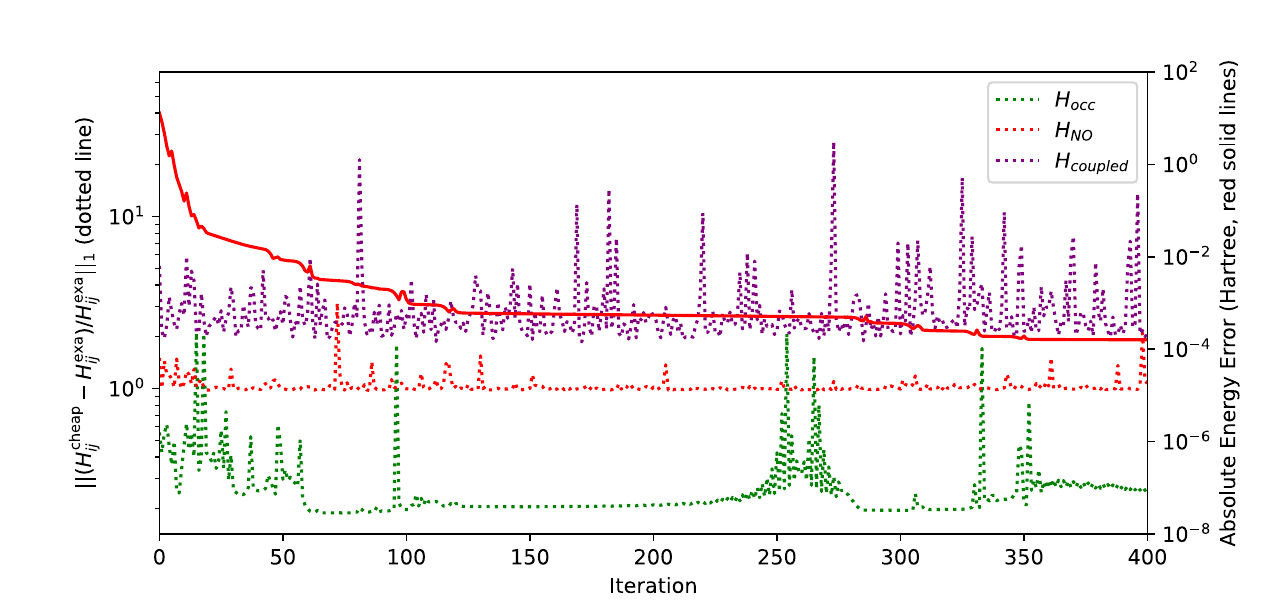}
    \end{minipage}
        
	\caption{ Relative error of $H^{\cheap}$ in the first iteration $\abs{\frac{H^{\cheap(0)}_{ij}-H^{\exa(0)}_{ij}}{H^{\exa(0)}_{ij}}}$ (top panels), the black lines indicate the separation between the different blocks of the hessian. And 1-norm of that relative error along the optimisation $\norm{\frac{ H^{\cheap}_{ij}-H^{\exa}_{ij}}{H^{\exa}_{ij}}}_1$ divided by the number of entries of the hessian block (dotted lines in bottom panels) for the ON block (green lines), NO block (red lines) and coupling block (purple lines). Results are shown for the \ce{H2O} (left panels) and \ce{C2H6} molecule (right panels).}
	\label{fig:hess_cheap_error}
\end{figure*}

To assess this assumption, we looked at the error made by taking $B=H^{\cheap}$ in Fig.~\ref{fig:hess_cheap_error} for the different blocks of the hessian. This shows that the ON block of the hessian is decently approximated by $H^{\cheap}$ while the off-diagonal elements of the NO block and the coupling block are less accurate. The pattern of the upper panels of Fig.~\ref{fig:hess_cheap_error} shows that the error in the NO block remains low when two indices of $X_{ij}$ and $X_{pq}$ are equal, that is, when $H^{\cheap}_{ijpq}\neq0$, highlighting the use of $H^{\cheap}$.

To avoid convergence issues for most systems, we thus need a suitable approximation of $H^{\exp}$ and will propose one in section \ref{sec:approx_hess}.

\section{Approximate hessians}
\label{sec:approx_hess}

The most common numerical approximation of the hessian in the literature is the BFGS approximation.\cite{NocedalWright2006}
The BFGS update is obtained by demanding that the new approximate hessian $B$ is close to the previous one $B^{(k)}$ under the constraints that the new approximate is symmetric and that $B$ satisfies the secant equation.\cite{NocedalWright2006} That is, the BFGS update solves
\begin{multline}\label{eq:BFGS_prob}
    \min_{B\in \mathds{R}^{N\times N}} \norm{ B-B^{(k)}} \\
    \quad\text{ s.t. } \quad B^T = B \text{ and } B s^{(k)} = y^{(k)},
\end{multline}
where $s^{(k)}$ and $y^{(k)}$ the step and difference of the gradient between the $(k+1)^{th}$ and $k^{th}$ iterations respectively. The constraint \(B s^{(k)} = y^{(k)}\) is the secant equation, which simply means that the hessian corresponds to a quadratic model through the last two iterates. Using the Frobenius norm, we obtain the standard BFGS update (see section~\ref{sec:appendix_BFGS} of the supplementary material for details)
\begin{equation}\label{eq:BFGS}
    B^{(k+1)} = B^{(k)} - \frac{B^{(k)} s^{(k)} s^{(k)T}B^{(k)} }{s^{(k)T} B^{(k)} s^{(k)}} + \frac{y^{(k)} y^{(k)T}}{s^{(k)T} y^{(k)}}.
\end{equation}
Unfortunately the bare BFGS approximation of the hessian does not provide a decent approximation of the exact hessian, leading to a poor convergence of the energy (see section~\ref{sec:BFGSconcergence} of the supplementary material).

The BFGS approximation is intended to approximate the hessian for convex problems.
However, due to the parametrization, we do the energy minimisation w.r.t.\ $\x = (x,X^{\text{up}})$, where $X^{\text{up}}$ is the vector of the strictly upper triangle entries of $X$. We will refer to the space of these variables as the $\x$-space.

On the other hand, we will call the $\NU$-space (`NU-space'), the space where the energy derivatives can be directly calculated, i.e.\ \(\NU = (\sqrt{n}, U)\), where $U$ comprises all entries of the full $N\times N$ matrix. We can readily transform the relevant derivatives from the $\NU$- to the $\x$-space using the Jacobian matrix $J_{pq} = \deriv{\NU_p}{\x_q}$ and its derivative.
The hessian tends to have a nicer expression in $\NU$-space (is often quartic in $U$ and the square roots of ONs) meaning that we can hope that the BFGS provides a better approximation in this space.
Also the fact that the unitary parametrization tends to make the hessian indefinite, indicates that the BFGS update on the hessian in $\NU$-space might work better.

Since we can readily calculate the cheap part of the hessian in $\Order(N^4)$ (see Sec.~\ref{sec:cheap_hess}), we only need to approximate the remaining part in $\NU$-space as $B^{\exp}_{\NU}\approx H^{\exp}_{\NU}$.

Denoting $\grad_{\NU}$ the gradient in $\NU$-space, $\grad^2_{\x}$ the hessian operator in $\x$-space, the expensive energy hessians $H^{\exp}$ in $\x$- and $\NU$-space are related by the chain rule as 
\begin{equation}\label{eq:hess_x_to_NU}
    H_{\x}^{\exp} = J^T H_{\NU}^{\exp} J + \grad_{\NU}^TE \,\grad^2_{\x}\NU,
\end{equation}
and the total hessian is 
\begin{equation}
    H_{\x} = H^{\cheap}_{\x} + J^T H_{\NU}^{\exp} J.
\end{equation}
Now we have several choices to define closeness to the previous step. One sensible option (another is presented in the supplementary material in Sec.~\ref{sec:Bexp-xclose}) is to make the hessians close in $\NU$-space, whilst still demanding the secant equation in $\x$-space, so
\begin{multline}
\min_{B}\norm\big{B - B_{\NU}^{\exp(k)}}  \\
\text{s.t. \quad 
\(B^T = B\) and \(B\bar{s}^{(k)}_{\NU} = \bar{y}^{(k)}_{\NU} - \bar{\xi}^{(k)}_{\NU}\)},
\end{multline}
where 
\begin{subequations}
\begin{align}\label{eq:def_barxi}
\bar{s}^{(k)}_{\NU} &\isDefinedAs J^{(k+1)}s_{\x}^{(k)} , \\
\bar{y}^{(k)}_{\NU} &\isDefinedAs J^{(k+1)-T}y_{\x}^{(k)},\\
\Bar{\xi}^{(k)}_{\NU} &\isDefinedAs J^{(k+1)-T}H_{\x}^{\cheap(k+1)}s_{\x}^{(k)}.
\end{align}
\end{subequations}

We thus only have to replace $B\rightarrow B_{\NU}^{\exp}$, $s\rightarrow \Bar{s}_{\NU}$ and $y\rightarrow \Bar{y}_{\NU}-\Bar{\xi}_{\NU}$ in~\eqref{eq:BFGS} and obtain
\begin{align}\label{eq:nuBFGS_aux_x}
B^{\exp(k+1)}_{\NU} &= B^{\exp(k)}_{\NU} - 
\frac{B^{\exp(k)}_{\NU} \bar{s}_{\NU}^{(k)} \bar{s}_{\NU}^{(k)T} B^{\exp(k)}_{\NU}}{\bar{s}_{\NU}^{(k)T} B^{\exp(k)}_{\NU} \bar{s}_{\NU}^{(k)}} \notag\\*
&+ \frac{\bigl(\bar{y}^{(k)}_{\NU} - \bar{\xi}^{(k)}_{\NU})\bigl(\bar{y}^{(k)}_{\NU} - \bar{\xi}^{(k)}_{\NU}\bigr)^T}{\bar{s}_{\NU}^{(k)T}\bigl(\bar{y}^{(k)}_{\NU} - \bar{\xi}^{(k)}_{\NU}\bigr)} .
\end{align}

\begin{figure*}[t]
        \includegraphics[width=\figsize]{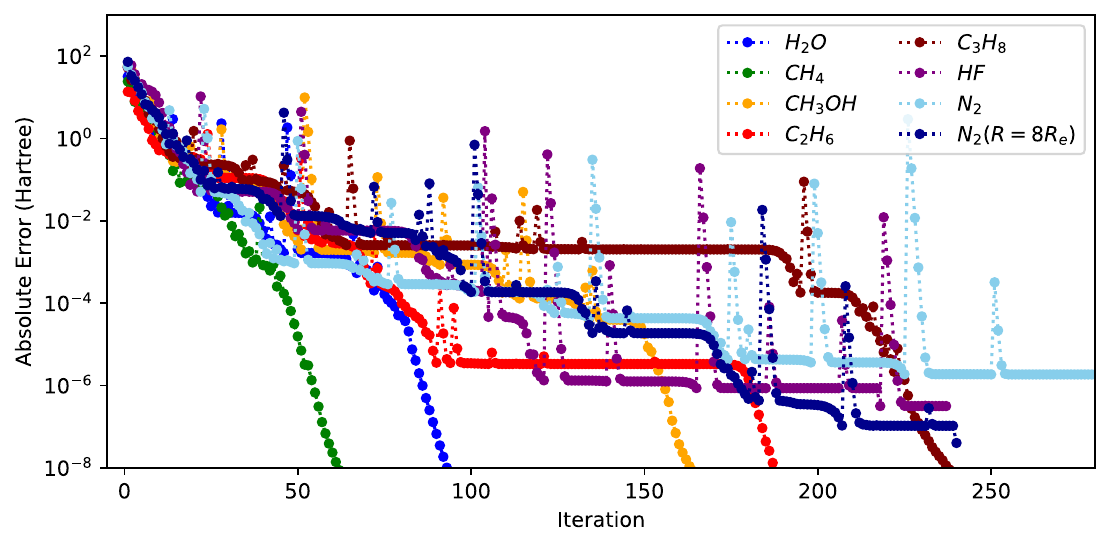}
	\caption{Convergence of the energy ($E^{(k)}-E^{(\text{ref})}$ with respect to micro-iteration count) using $H^{\cheap}_{\x}$ exactly and equation~\eqref{eq:nuBFGS_aux_x} to approximate $H^{\exp}$. }
	\label{fig:no_damping_conv}
\end{figure*}

This approximation 
(Fig.~\ref{fig:no_damping_conv}) unfortunately cannot match the good behaviour of the $B_{\x}=H^{\cheap}_{\x}$ approximation (Fig.~\ref{fig:cheap_hess_conv}) for the first iterations. 
In section~\ref{sec:cheap_hess} we attributed the initial good convergence of the $B_{\x}=H^{\cheap}_{\x}$ approximation to a sufficient approximation of the hessian.
To take full advantage of the good convergence of $B_{\x}^{\exp}=0$, we then decided to turn on the BFGS approximation of $H^{\exp}$ only when the algorithm start to converge with $B_{\x}=H^{\cheap}_{\x}$, to correct the convergence if it was not going to an actual minimum. Since $H^{\cheap}_{\x}$ approximated the ON block of the hessian particularly well, it seems reasonable to look primarily at the ON step. In practice, we add a factor $A(s_n,\epsilon)=\frac{1}{2}\bigl[1-\erf\bigl(\log_{10}(\abs{s_n}_{\infty}/\epsilon)\bigr)\bigr]$ to $B_{\x}^{\exp}$, where $|s_n|_{\infty}$ is the $\infty$-norm of the ON step.

\begin{figure*}[t]
        \includegraphics[width=\figsize]{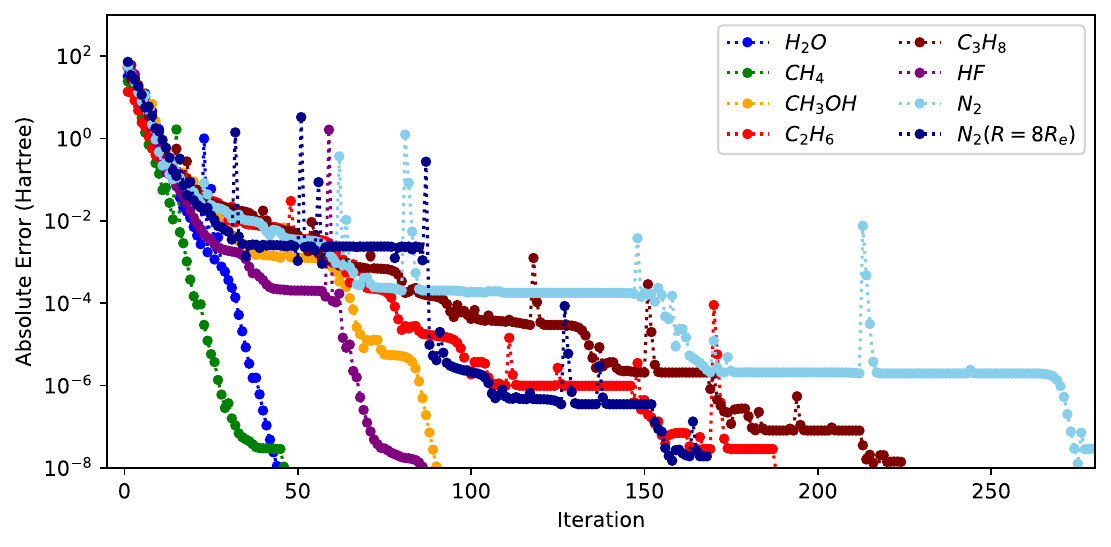}
	\caption{Convergence of the energy ($E^{(k)}-E^{(\text{ref})}$ with respect to micro-iteration count) using $H^{\cheap}_{\x}$ exactly and equation~\eqref{eq:nuBFGS_aux_x} to approximate $H_{\x}^{\exp}$ with the prefactor $A(s_n,10^{-3})$ (see text). }
	\label{fig:mixed_bhess_conv}
\end{figure*}

We report the result in Fig.~\ref{fig:mixed_bhess_conv}. By comparing Fig.~\ref{fig:no_damping_conv} and Fig.~\ref{fig:mixed_bhess_conv} we see that the prefactor $A$ significantly improves the convergence, even beyond the first few iterations. Although the convergence is not as good as for the exact hessian (Fig.~\ref{fig:exa_hess_conv}), it is able to reach an error of $2\cdot10^{-8}$ Hartree within 280 iterations for all molecules. To reach $10^{-8}$ Hartree, we would need to use a stronger termination condition than our default settings (see Fig.~\ref{fig:1step_vs_2step_algo}). 

\begin{figure*}[tp]
        \includegraphics[width=\figsize]{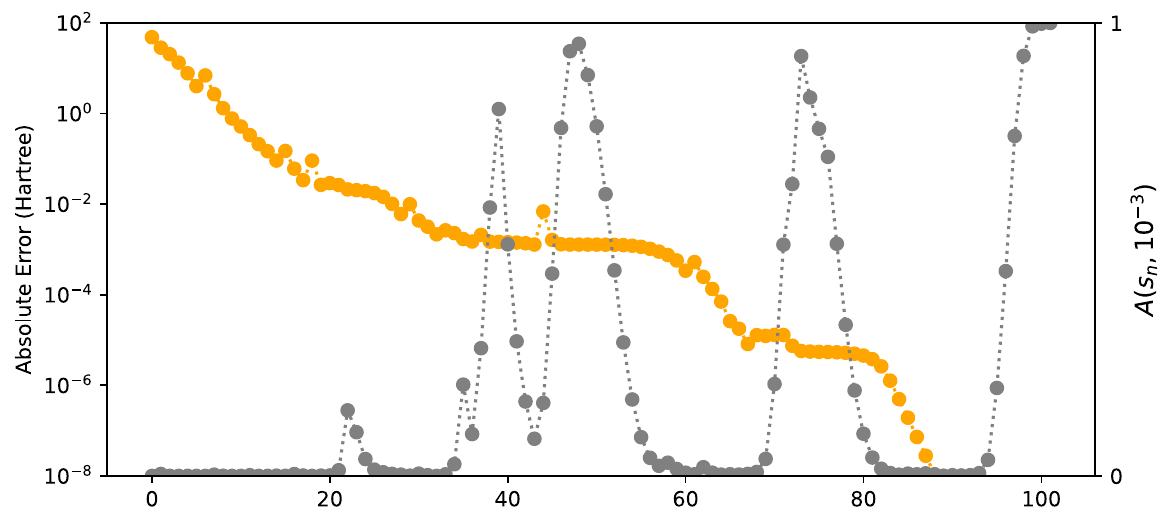}
	\caption{Absolute energy error (orange line) and the prefactor $A(s_n,10^{-3})$ used on $B^{\exp}$ in \mbox{$\NU$-space} (grey line) w.r.t.\ the iteration during the SCF procedure, using equation~\eqref{eq:nuBFGS_aux_x},for the \ce{CH3OH} molecule. }
	\label{fig:prefactor_behaviour}
\end{figure*}

It may be informative to look at the evolution of $A$ to understand the plateaus in the convergence of the energy for some molecules using $B^{\exp}_{\NU}$. We plot $A$ and the energy error as functions of the iteration in Fig.~\ref{fig:prefactor_behaviour}. Only the result for the \ce{CH3OH} molecule is reported here, but the results are qualitatively similar for the other molecules that do not converge with the $B^{\exp}=0$ approximation. We observe that $A$ does go from 0 to 1, but oscillates, inducing the plateaus in Fig.~\ref{fig:mixed_bhess_conv}. This indicates that the choice of $A$ is not optimal and that the algorithm can be improved by building a better prefactor.

In order to compare the above approximation with existing results, we tested our molecules with the Müller functional implemented in the Natural Orbital Functional (NOF) Theory module\cite{RodriguezMayorga2022} of the molgw software.\cite{BrunevalRangel2016} This module is based on the widely use RDMFT program, DoNOF.\cite{PirisMitxelena2021} DoNOF was developed for recent Piris NOF functionals\cite{PirisLopez2011,Piris2017,Piris2021} and inherits the partitioning of the orbitals in seniority-zero subspaces, meaning that the `Müller' functional in this module does have to comply with additional constraints, but is believed to by decently close to the actual Müller functional. 

\begin{table}[tb]
    \footnotesize
    \centering
    \caption{\normalsize Number of iterations for the molgw implementation\cite{BrunevalRangel2016,RodriguezMayorga2022} of the DoNOF code\cite{PirisMitxelena2021} and our last approximation (Fig.~\ref{fig:mixed_bhess_conv}) to converge to an energy error of $10^{-3}$ Hartree for different molecules (the reference energy for the DoNOF code is the energy obtained after a large number of 200 to 300 iterations and for our code the energy obtained with the exact hessian in fig.~\ref{fig:exa_hess_conv}). For DoNOF we report the number of macro-iterations (left column), the total number of NO iteration (middle column) and the total number of ON iterations (right column).}
    \label{tab:iteration_DoNOF_vs_bBFGS}
    \begin{tabular}{c|c|c|c|c}
        \toprule
        {} & \multicolumn{3}{|c|}{DoNOF} & 
        Fig.~\ref{fig:mixed_bhess_conv}\\
         molecule   & macro-it. & NO it. & ON it. & it. \\
         \midrule
         \ce{H2O}   &  96 & 2851 &  2266 &  28 \\
         \ce{CH4}   & 104 & 3091 &  3186 &  19 \\
         \ce{CH3OH} & 185 & 5521 &  8035 &  59 \\
         \ce{C2H6}  & 181 & 5401 & 17553 &  62 \\
         \ce{HF}    & 219 & 6541 & 12581 &  36 \\
         \ce{N2}    & 150 & 4471 & 13424 &  61 \\
         \ce{N2}$(8R_e)$    & 170 & 5071 & 13123 &  87 \\ 
         \bottomrule
    \end{tabular}
\end{table}

More precisely DoNOF employes a 2-step procedure, optimising the ONs with a limited memory BFGS approximation and the NOs by using an iterative diagonalisation approach (with a maximum of 30 NO iteration per macro-iteration). The later corresponds to a steepest descent and thus struggles to converge to high precision. For that reasons, it has been decided report the results to reach an error of $10^{-3}$ Hartree (instead of $10^{-8}$ up to now) in Table~\ref{tab:iteration_DoNOF_vs_bBFGS}. The DoNOF implementation does (exactly or with reduction of identity) the 4-index transformation explicitly, which needs to be computed at each NO iteration but only at the first ON iteration of a macro-iteration, meaning that the total number of expensive iterations is equal to the number NO iterations plus the number of macro-iterations. On the contrary if it would use the idea of ref~\cite{Giesbertz2016}, to avoid explicit 4-index transformations, the cost of NO and ON iterations would be of the same order and the total number of expensive iterations the sum of the NO and ON iterations. We thus report in table~\ref{tab:iteration_DoNOF_vs_bBFGS}, the number of macro-, NO and ON iterations. 
To compare to our code, we will focus on the number of expensive iterations as implemented in DoNOF (i.e.\ macro- plus NO iterations). We obtain that DoNOF needs between 3000 and 7000 iterations while our algorithm always reaches the required error within 30 to 90 iterations, which is an improvement of two orders of magnitude.

\section{Conclusion}
\label{sec:conclusion}

In this work, we used an optimisation based on the hessian to reduce the number of iterations of the SCF-RDMFT procedure. We first added the cross terms in the hessian, providing an efficient 1-step optimisation method, in contrast to the more common 2-step one. As a baseline, we have shown that a trust-region optimization with the exact hessian provides a very effective algorithm for this. However, using the exact hessian, makes the iterations too expensive to be useful in practice. Thus we have derived an accurate approximation of the hessian. We first extracted an affordable part of the exact hessian. Using only this cheap part, the number of iterations is actually of similar order as for the exact hessian if it converged, but half of our test set actually did not converge. A rather robust and effective approximation was obtained by using the cheap part of the hessian exactly and use a BFGS-like approximation to only approximate the expensive part in the more suitable $\NU$-space. This approximate hessian resulted in an algorithm which converged for all eight molecules within a few hundred iterations to an error less than $10^{-8}$ Hartree.

Although the numerical results show a great amelioration of the convergence compared to a naive algorithm, there is still some room of improvement compared to the exact hessian. To enhance the convergence even further, a more accurate approximation of the hessian seems to be required. A first step in this direction would be to build a consistent secant equation in $\x$-space. 

The plateaus in the energy versus iteration probably indicate that the algorithm is struggling with relatively flat areas in the energy landscape i.e.\ NOs close to 0 or 2, which may become especially problematic for large basis sets. This might also explain why when using even the exact hessian, convergence is not that fast as one would expect. So another direction to improve the algorithm is to avoid these flat areas, by modifying the parametrisation. 

We have only used the Müller functional in our tests. Although the the approximate hessian presented here can be applied to non-separable functionals, the computation of the energy for these functionals is already in $\Order(N^5)$ and using $H^{\exa}$ would therefore be preferable. Moreover we can use the resolution of identity to obtain a scaling of $\Order(N^4)$ for both the computation of the energy and $H^{\exa}$. However non-separable RDMFT functionals are not convex and some even display discontinuities respect to the occupations, which may hinder the convergence. The efficiency of our approach for non-separable functionals has thus yet to be assessed, which will be subject of future work.


\begin{acknowledgement}
The authors thank the The Netherlands Organisation for Scientific Research, NWO, for its financial support under Grant No.~OCENW.KLEIN.434 and Vici Grant No.~724.017.001.
\end{acknowledgement}


\appendix

\section{Expression of the exact hessian}
\label{sec:appendix_Hexa}

The derivatives of the 1-RDM w.r.t.\ to the $x_i$ are obtained by simply taking the derivative of the NOs in~\eqref{eq:1RDM_0th}. Likewise, we get the cross-derivative of the 1-RDM by taking the derivative of the NOs w.r.t.\ the $x_i$ in~\eqref{eq:dgammadX}.\\
Inserting the derivatives of the 1-RDM in~\eqref{eq:E_split} we get, for the NO block of the hessian
\begin{align}
    \label{eq:ddE1ddx}
    \left.\derivw{E_1}{x_i}{x_j}\right\vert_0 &= \sum_p \derivw{n_p}{x_i}{x_j}H^1_{pp} \\
    \label{eq:ddEHddx}
    \left.\derivw{E_H}{x_i}{x_j}\right\vert_0 &=\sum_{pq} \left( \derivw{n_p}{x_i}{x_j} n_q +\deriv{n_p}{x_i}\deriv{n_q}{x_j}\right)[pp|qq] \\
    \label{eq:ddExcddx}
    \left.\derivw{W_{\xc}}{x_i}{x_j}\right\vert_0 &= \sum_{pq}\derivw{F(n_p,n_q)}{x_i}{x_j}[pq|pq],
\end{align}
for the coupling block, 
\begin{align}
    \label{eq:ddE1dXdx}
    \left.\derivw{E_1}{X_{ij}}{x_k}\right\vert_0 &= 2\left(\deriv{n_i}{x_k} - \deriv{n_j}{x_k} \right)H^1_{ij}\\
    \label{eq:ddEHdXdx}
    \left.\derivw{E_H}{X_{ij}}{x_k}\right\vert_0 &= 2\sum_p \bigg((n_i-n_j)\deriv{n_p}{x_k} \notag \\ 
    &+ (\deriv{n_i}{x_k} - \deriv{n_j}{x_k})n_p\bigg)[ij|pp]  \\
    \label{eq:ddExcdXdx}
    \left.\derivw{W_{\xc}}{X_{ij}}{x_k}\right\vert_0 &= \sum_p\left(\deriv{F(n_i,n_p)}{x_k}-\deriv{F(n_j,n_p)}{x_k}\right)[ip|jp] \notag \\
    &+ \left(\deriv{F(n_p,n_i)}{x_k}-\deriv{F(n_p,n_j)}{x_k}\right)[jp|ip] \\
            &=2\sum_p\left(\deriv{F(n_i,n_p)}{x_k}-\deriv{F(n_j,n_p)}{x_k}\right)[ip|jp],
\end{align}
if $F(n_i,n_j) = F(n_j,n_i)$, which holds for all functionals we are aware of. And, for the ON block, we get
\begin{align}
    \label{eq:ddE1ddX}
        \left.\derivw{E_1}{X_{ij}}{X_{pq}}\right\vert_0 &=
             \delta_{ip}(2n_i - n_j - n_q)H^1_{jq}  \notag\\
            &+ \delta_{jq}(2n_j - n_i - n_p)H^1_{ip}\notag\\
            &- \delta_{iq}(2n_i - n_j - n_p)H^1_{jp}\notag\\
	    &- \delta_{jp}(2n_j - n_i - n_q)H^1_{iq},
\end{align}
\begin{align}
    \label{eq:ddEHddX}
        \left.\derivw{E_{H}}{X_{ij}}{X_{pq}}\right\vert_0 =
            \sum_k &\Big( \hspace{8pt} 2\delta_{pi}(2 n_i n_k- n_j n_k- n_q n_k)[qj|kk]\notag\\
		&+ 2\delta_{qj} (2 n_j n_k - n_i n_k - n_p n_k)[ip|kk]\notag \\
		&- 2\delta_{iq}	(2 n_i n_k - n_j n_k - n_p n_k)[jp|kk]\notag \\
		&- 2\delta_{jp} (2 n_j n_k - n_i n_k - n_q n_k)[iq|kk] \Big)\notag \\
		&+ 4(n_i n_p - n_j n_p - n_i n_q + n_j n_q)[ij|pq],
\end{align}
\begin{align}
    \label{eq:ddExcddX}
        \left.\derivw{W_{\xc}}{X_{ij}}{X_{pq}}\right\vert_0 =
            \sum_k &\Big( \hspace{8pt} 2\delta_{pi}(F(n_i,n_k)-F(n_j,n_k) \notag \\
            & + F(n_p,n_k) -F(n_q,n_k))[qk|jk] \notag \\
		&+ 2\delta_{qj} (F(n_j,n_k) -F(n_i,n_k) \notag \\
            & + F(n_q,n_k) -F(n_p,n_k))[ik|pk] \notag \\
		&- 2\delta_{iq}	(F(n_i,n_k)-F(n_j,n_k) \notag \\
            & + F(n_q,n_k) -F(n_p,n_k))[jk|pk] \notag \\
		&- 2\delta_{jp} (2F(n_j,n_k)-F(n_i,n_k) \notag \\
            & + F(n_p,n_k) -F(n_q,n_k))[ik|qk] \Big) \notag \\
		&+ 4\big(F(n_i,n_p) -F(n_j,n_p) \notag \\
            & -F(n_i,n_q) + F(n_j,n_q)\big)[ip|jq].	   
\end{align}

\newpage
\bibliography{bibliography.bib}

\addtocounter{figure}{-1}
\refstepcounter{figure}\label{LASTFIGURE}
\addtocounter{equation}{-1}
\refstepcounter{equation}\label{LASTEQ}

\end{document}


\section{Relevance of the coupling block of the approximate \hessian }
\label{sec:couplingApproxHes}

\begin{figure*}[t!]
        \centering
        \includegraphics[width=\figsize]{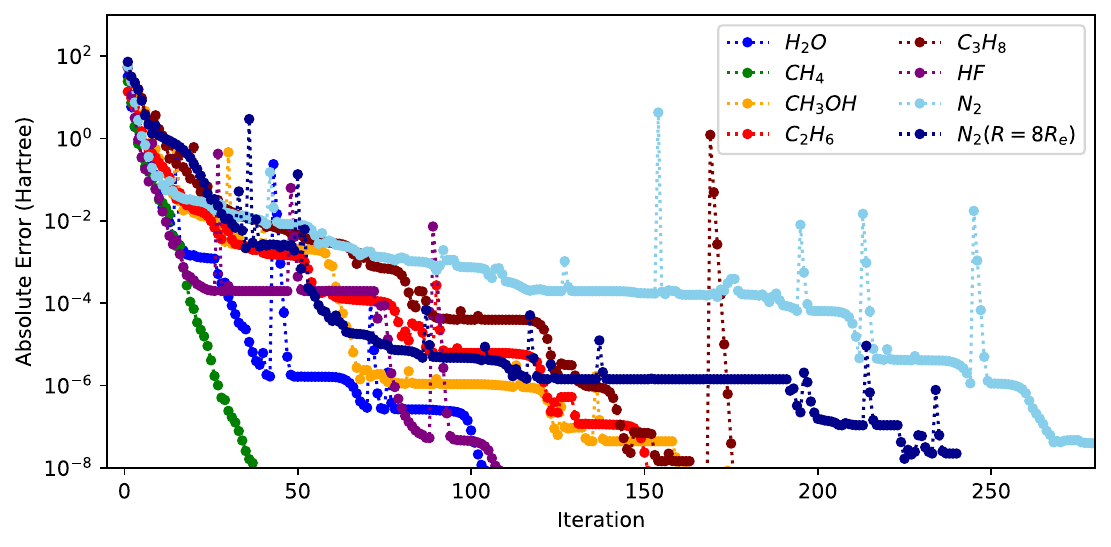}
	\caption{Convergence of the energy ($E^{(k)}-E^{(\text{ref})}$ with respect to \microiteration count) using $H^{\cheap}_{\x}$ exactly and~\eqref{eq:nuBFGS_aux_x} to approximate $H_{\x}^{\exp}$ with the prefactor $A(s_n,10^{-3})$ (see text in the article) for the ON-ON and NO-NO blocks of the \hessian and setting the ON-NO block to 0 for different molecules. }
	\label{fig:bBFGS_no_coupling}
\end{figure*}

In the article we compared the convergence with and without coupling block for the exact newton method (\Section~\ref{sec:1_step}), and concluded that the coupling block did not significantly improved the convergence rate in that case: for some systems the energy converged faster, others slower.
A more consistent improvement of the convergence rate is actually visible with an approximate \hessian. Here we tested the approximation~\eqref{eq:nuBFGS_aux_x}, and can indeed see by comparing \Fig~\ref{fig:bBFGS_no_coupling} with~\ref{fig:mixed_bhess_conv} a faster convergence when adding the (approximate) coupling block. The plots not only display less plateaus, but it also takes less iterations for the algorithm to exit them (see \ce{H2O}, \ce{CH3OH}, \ce{HF} and stretched \ce{N2}).  
A possible explanation is that the coupling block may help to determine early in the convergence which orbitals are almost fully or not occupied and avoid steps that would tend to make ONs too close to 0 and 2.

\section{Convergence of the BFGS approximation}
\label{sec:BFGSconcergence}

\begin{figure*}[t!]
        \centering
        \includegraphics[width=\figsize]{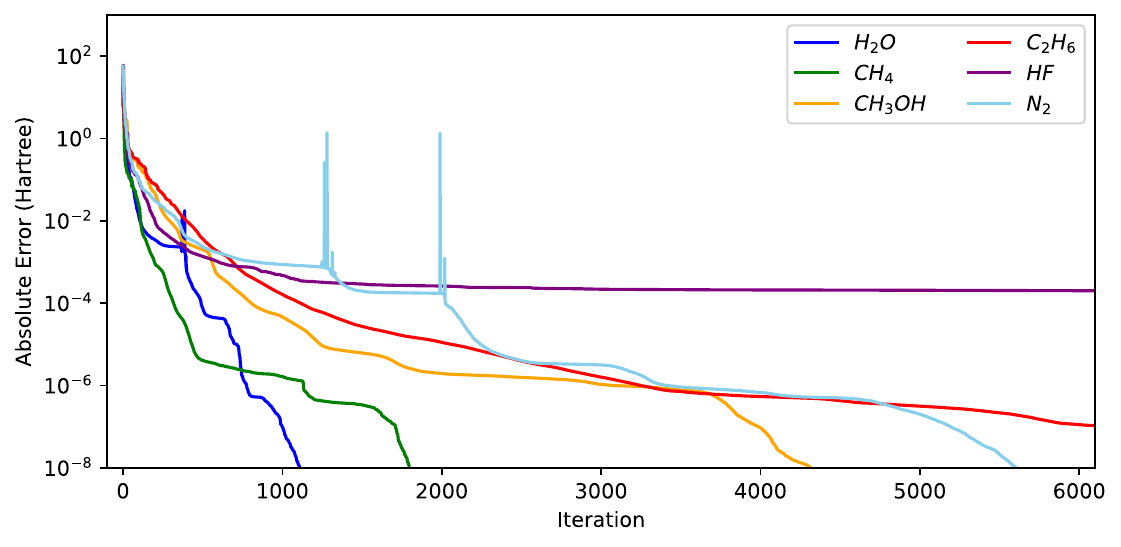}
	\caption{Convergence of the energy ($E^{(k)}-E^{(\text{ref})}$ with respect to \microiteration count) using a 1-step procedure with \hessian approximated by the BFGS method for different molecules. }
	\label{fig:BFGS_conv}
\end{figure*}

\begin{figure*}[t!]
        \centering
        \includegraphics[width=\figsize]{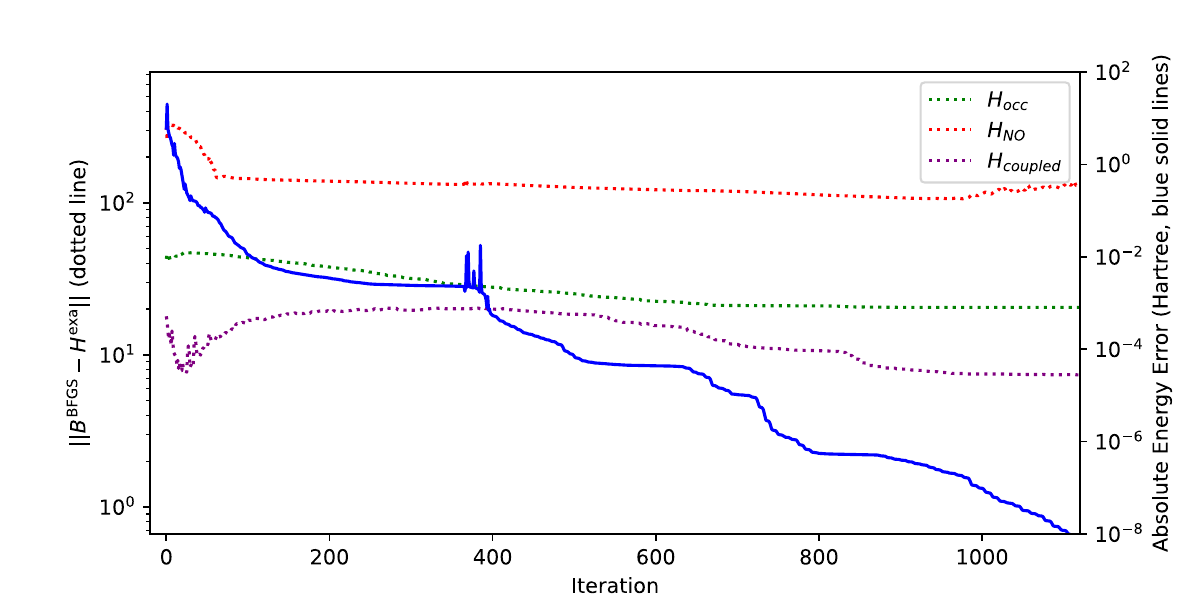}
        \includegraphics[width=\figsize]{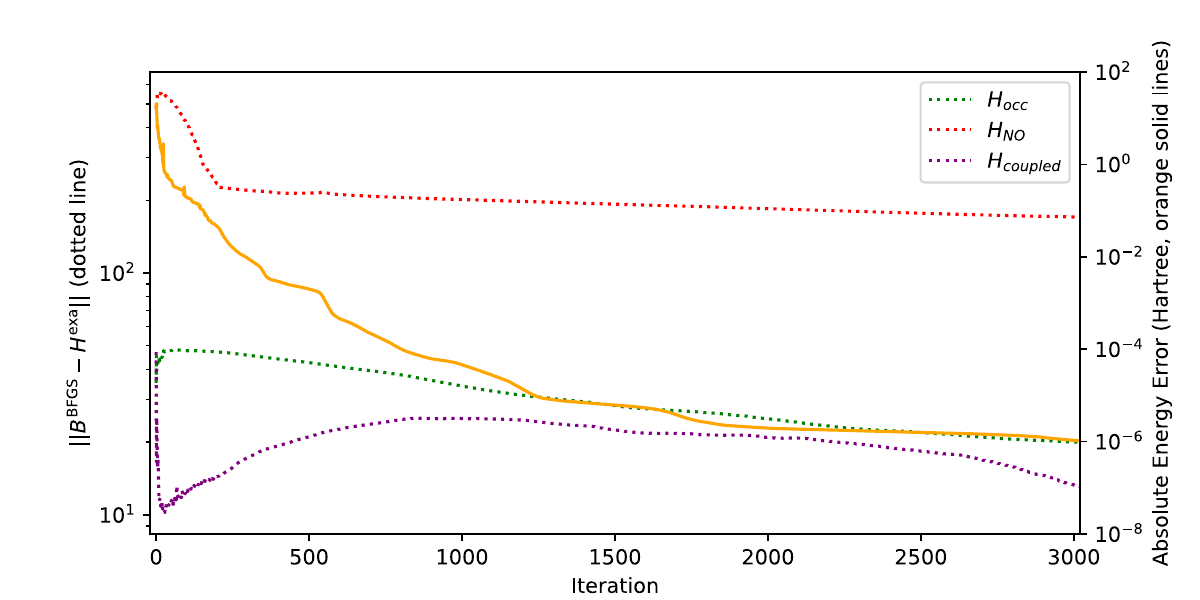}
	\caption{Error of the BFGS approximation $\lVert B^{\text{BFGS}}-H^{\exa}\rVert$ (dotted line) 
    for the \ce{H2O} molecule (top) and \ce{CH3OH} molecule (bottom). The error coming from the ON block is plotted as dotted red lines, from the NO block as green lines and from the coupling blocks as purple lines.}
	\label{fig:BFGS_hess_error}
\end{figure*}

The BFGS approximation itself, given by~\eqref{eq:BFGS}, (see Sec.~\ref{sec:appendix_BFGS} for the derivation) does not provide a good convergence of the energy, as illustrated in \Fig~\ref{fig:BFGS_conv}. 
The error induced by the BFGS approximation is actually large, as can be seen from \Fig~\ref{fig:BFGS_hess_error}, where we plotted the error made by the BFGS approximation compared to the exact \hessian. Since the BFGS approximation encourages a positive \hessian it has a hard time reproducing correctly the exact \hessian (which is indefinite, see \Fig~\ref{fig:exa_hess_neigvals}) for the first few iterations, but is supposed to reduce the error while getting closer to the minimum. However, we observe that this is not the case, indicating that the BFGS approximation does not suffice to efficiently approximate the exact hessian. This motivated us to obtain a better approximation of the \hessian in the article.

\section{Alternative BFGS approximations}
\subsection{Approximation~\eqref{eq:nuBFGS_aux_x} for the total \hessian}
In the article we investigated a modified BFGS approximation~\eqref{eq:nuBFGS_aux_x}, applied to the expensive part of the \hessian $H^{\exp}$ only. Note that a similar approximation can be derived for the total \hessian too. 

Defining 
\begin{align}
\bar{\rho}^{(k)}_{\NU} &\isDefinedAs J^{(k+1)-T}\grad^T_{\NU}E\vert_{(k+1)}\grad^2_{\x} \NU^{(k+1)}s^{(k)}_{\x}.
\end{align}
We simply need to set \(y^{(k)} \to \bar{y}^{(k)}_{\NU} - \bar{\rho}^{(k)}_{\NU}\), \(B^{(k)} \to B_{\NU}^{(k)}\) and \(s^{(k)} \to \bar{s}^{(k)}_{\NU} \) in~\eqref{eq:BFGS}, which yields
\begin{subequations}\label{eq:BFGS_closeNU_secantX}
\begin{align}\label{eq:BFGS_closeNU_secantX_inNU}
B^{(k+1)}_{\NU} &= B^{(k)}_{\NU} - 
\frac{B^{(k)}_{\NU} \bar{s}_{\NU}^{(k)} \bar{s}_{\NU}^{(k)T} B^{(k)}_{\NU}}{\bar{s}_{\NU}^{(k)T} B^{(k)}_{\NU} \bar{s}_{\NU}^{(k)}} \notag\\*
&+ 
\frac{\bigl(\bar{y}^{(k)}_{\NU} - \bar{\rho}^{(k)}_{\NU})\bigl(\bar{y}^{(k)}_{\NU} - \bar{\rho}^{(k)}_{\NU}\bigr)^T}{\bar{s}_{\NU}^{(k)T}\bigl(\bar{y}^{(k)}_{\NU} - \bar{\rho}^{(k)}_{\NU}\bigr)} , \\
\intertext{and transformed to $\x$-space via~\eqref{eq:hess_x_to_NU} becomes}
B^{(k+1)}_{\x} &= \tilde{B}^{(k)}_{\x} - 
\frac{\Bar{B}^{(k)}_{\x} s_{\x}^{(k)} s_{\x}^{(k)T} \Bar{B}^{(k)}_{\x}}{s_{\x}^{(k)T} \Bar{B}^{(k)}_{\x} s_{\x}^{(k)}} \notag\\*
&+
\frac{\bigl(y_{\x}^{(k)} - \rho^{(k)}_{\x}\bigr)\bigl(y_{\x}^{(k)} - \rho^{(k)}_{\x}\bigr)^T}{s_{\x}^{(k)T}\bigl(y_{\x}^{(k)} - \rho^{(k)}_{\x}\bigr)} 
\end{align}
\end{subequations}
where \(\rho^{(k)}_{\x} \isDefinedAs J^{(k+1)T}\bar{\rho}^{(k)}_{\NU} = \grad^T_{\NU}E\vert_{(k+1)}\grad^2_{\x} \NU^{(k+1)}s^{(k)}_{\x}\), \(\Bar{B}^{(k)}_{\x} = J^{(k+1)T}B^{(k)}_{\NU}J^{(k+1)}\).

\subsection{BFGS with closeness in $\x$-space}\label{subsec:BFGS_close_X}

We focused, in the article on a modified BFGS approximation~\eqref{eq:nuBFGS_aux_x}, based on the closeness in $\NU$-space. An other option is to make the update close to the previous \hessian with updated Jacobian terms
\begin{align}\label{eq:BtildeDef}
\tilde{B}_{\x}^{(k)} \isDefinedAs J^{(k+1)T}B_{\NU}^{(k)}J^{(k+1)} + \grad^T_{\NU}E\vert_{(k+1)}\grad^2_{\x} \NU^{(k+1)}.
\end{align}
The BFGS optimization problem now becomes
\begin{multline}\label{eq:BFGS_x_aux_prob}
\min_{B}\norm\big{B - \tilde{B}_{\x}^{(k)}}  \quad \\
\text{s.t.} \quad 
\text{\(B^T = B\) and \(Bs_{\x}^{(k)} = y_{\x}^{(k)}\)} ,
\end{multline}
where $s_{\x}^{(k)}$, $y_{\x}^{(k)}$ are the step and difference of gradient between the $(k+1)^{th}$ and $k^{th}$ iterations in $\x$-space.
So we simply need to set \(y^{(k)} \to y_{\x}^{(k)}\), \(B^{(k)} \to \tilde{B}_{\x}^{(k)}\) and \(s^{(k)} \to s_{\x}^{(k)}\), which yields the update
\begin{subequations}\label{eq:BFGSaux_closeX_secantX}
\begin{align}
    B_{\x}^{(k+1)} &= \B_{\x}^{(k)} 
    - \frac{\B_{\x}^{(k)}s_{\x}^{(k)}s_{\x}^{(k)T}\B_{\x}^{(k)}}{s_{\x}^{(k)T}\B_{\x}^{(k)}s_{\x}^{(k)}} +\frac{y_{\x}^{(k)}y^{(k)T}_{\x}}{y^{(k)T}_{\x}s^{(k)}_{\x}} , \\
\intertext{which can be transformed to the update in $\NU$-space via the chain rule,}
    B_{\NU}^{(k+1)} &= B_{\NU}^{(k)} +\frac{\bar{y}^{(k)}_{\NU}\bar{y}^{(k)T}_{\NU}}{\bar{y}^{(k)T}_{\NU}\bar{s}^{(k)}_{\NU}}\notag\\*
    &- \frac{\bigl(B_{\NU}^{(k)}\bar{s}^{(k)}_{\NU} + \bar{\rho}^{(k)}_{\NU}\bigr)\bigl(B_{\NU}^{(k)}\bar{s}^{(k)}_{\NU} + \bar{\rho}^{(k)}_{\NU}\bigr)^T}
    {\bar{s}^{(k)T}_{\NU}\bigl(B_{\NU}^{(k)}\bar{s}^{(k)}_{\NU} + \bar{\rho}^{(k)}_{\NU}\bigr)}.
    \label{eq:BFGSaux_closeX_secantX_NUspace}
\end{align}
\end{subequations}

\begin{figure*}[t!]
        \centering
        \includegraphics[width=\figsize]{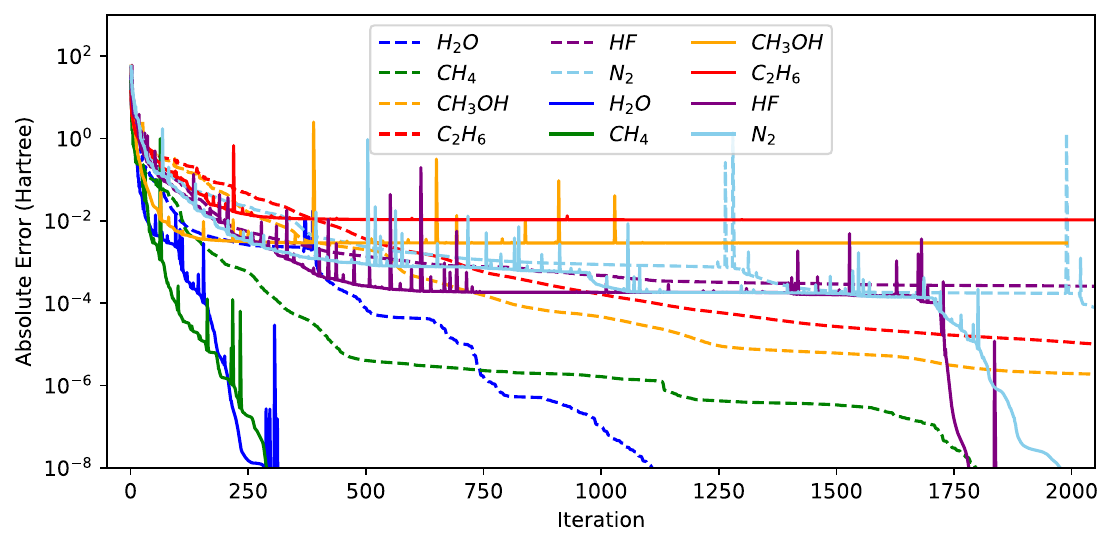}
	\caption{Convergence of the energy ($E^{(k)}-E^{(\text{ref})}$ with respect to \microiteration count) using equation~\eqref{eq:BFGSaux_closeX_secantX} in solid lines and~\eqref{eq:BFGS} in dashed lines for different molecules. }
	\label{fig:conv_H_aux_x}
\end{figure*}

The improvement w.r.t.\ the BFGS in $\x$-space (dashed lines) is significant for most molecules of the test set (see \Fig~\ref{fig:conv_H_aux_x}), but the algorithm shows difficulties to converge for the \ce{HF}, \ce{CH3OH} and \ce{C2H6} molecules. Indeed, it spends hundreds of iterations at an error of $2\cdot10^{-4}$ Hartree before resuming the convergence for the \ce{HF} molecule (purple line) and simply stops without finding the correct minimum for the other two molecules.

\subsection{$B^{\exp}$ approximation with closeness in $\x$-space}
\label{sec:Bexp-xclose}

As we restricted~\eqref{eq:BFGS_closeNU_secantX} to $B^{\exp}$ to obtain~\eqref{eq:nuBFGS_aux_x} in the article, we can make the updated \hessian close to the previous \hessian with updated Jacobian terms and \(H^{\cheap}\)
\begin{align}
\tilde{B}_{\x}^{\exp(k)} \isDefinedAs J^{(k+1)T}B_{\NU}^{\exp(k)}J^{(k+1)}+ H^{\cheap(k+1)}_{\x}
\end{align}
to restrict~\eqref{eq:BFGSaux_closeX_secantX_NUspace} to $B^{\exp}$.
We simply have to replace \(\tilde{B}_{\x}^{(k)} \to \tilde{B}_{\x}^{\exp(k)}\) in~\eqref{eq:BFGSaux_closeX_secantX_NUspace} and \(\bar{\rho}_{\NU}^{(k)} \to \bar{\xi}_{\NU}^{(k)}\) ($\bar{\xi}_{\NU}$ being defined in~\eqref{eq:def_barxi}), to get the update

\begin{align}\label{eq:xBFGS_aux_x}
    B_{\NU}^{\exp(k+1)} &= B_{\NU}^{\exp(k)} +\frac{\bar{y}^{(k)}_{\NU}\bar{y}^{(k)T}_{\NU}}{\bar{y}^{(k)T}_{\NU}\bar{s}^{(k)}_{\NU}} \notag\\*
    &- \frac{\bigl(B_{\NU}^{\exp(k)}\bar{s}^{(k)}_{\NU} + \bar{\xi}^{(k)}_{\NU}\bigr)\bigl(B_{\NU}^{\exp(k)}\bar{s}^{(k)}_{\NU} + \bar{\xi}^{(k)}_{\NU}\bigr)^T}
    {\bar{s}^{(k)T}_{\NU}\bigl(B_{\NU}^{\exp(k)}\bar{s}^{(k)}_{\NU} + \bar{\xi}^{(k)}_{\NU}\bigr)}.
\end{align}

\begin{figure*}[t!]
        \centering
        \includegraphics[width=\figsize]{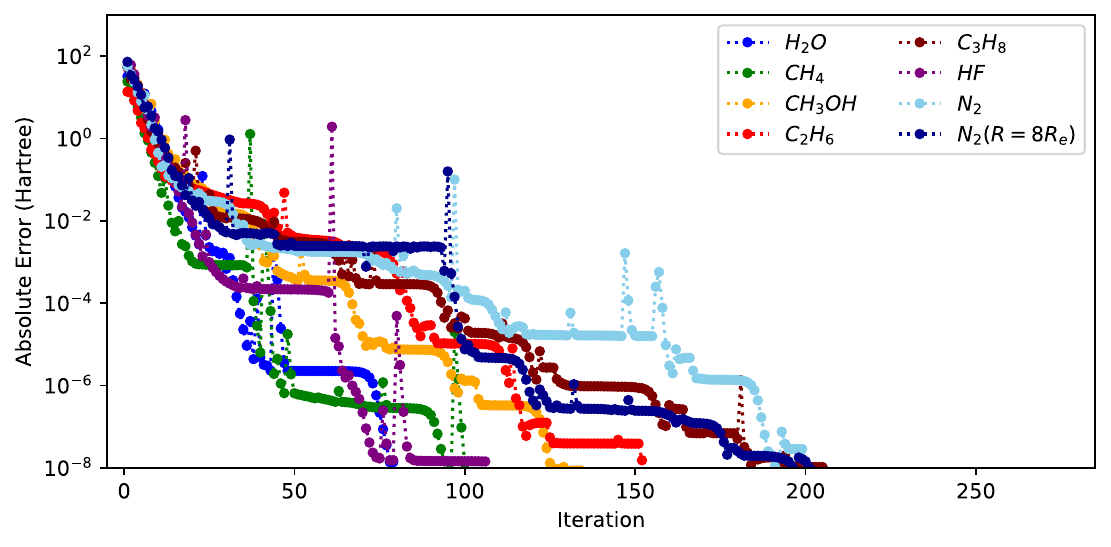}
	\caption{Convergence of the energy ($E^{(k)}-E^{(\text{ref})}$ with respect to \microiteration count) using $H^{\cheap}_{\x}$ exactly and~\eqref{eq:xBFGS_aux_x} to approximate $H_{\x}^{\exp}$ with the prefactor $A(s_n,10^{-3})$ (see text in the article). }
	\label{fig:mixed_thess_conv}
\end{figure*}

We plot the convergence using this approximation, with a damping prefactor $A$, as described in \Section~\ref{sec:approx_hess} of the article, see \Fig~\ref{fig:mixed_thess_conv}.
The update~\eqref{eq:xBFGS_aux_x} seems to be overall less efficient than~\eqref{eq:nuBFGS_aux_x}, especially for small molecules (\ce{H2O}, \ce{CH4} and \ce{HF}), 
suggesting that the prefactor should be modified for this approximation. Yet,~\eqref{eq:nuBFGS_aux_x} is not systematically better (compare the convergences of \ce{N2} and \ce{C2H6} between \Fig~\ref{fig:mixed_bhess_conv} and~\ref{fig:mixed_thess_conv}).

\subsection{BFGS with the secant equation in $\NU$-space}\label{subsec:BFGS_seceq_NU}

It needs to be mentioned that we do not implement the gradient difference exactly as needed in BFGS-like updates (\eqref{eq:nuBFGS_aux_x} and~\eqref{eq:BFGSaux_closeX_secantX}), since it reads in full
\begin{align}
y_{\x}^{(k)} &= J^{(k+1)}(0)\grad_{\NU}E\bigr\rvert_{(k+1)} \notag\\* 
             &- J^{(k+1)}\bigl(-s_{\x}^{(k)}\bigr)\grad_{\NU}E\bigr\rvert_{(k)} .
\end{align}
Indeed, the Jacobian is very tough to evaluate for the exponential parametrization for $X \neq 0$. As a pragmatic solution, assuming that the NO step, $s_{NO}$ is small enough (in $\x$-space), we only take the zeroth order term from its Taylor expansion for the part parametrizing the NOs,
\begin{align}\label{eq:jac_0th}
J^{(k+1)}(-\{s_{n\x}^{(k)},s_{NO\x}^{(k)}\}) \approx J^{(k+1)}(-\{s_{n\x}^{(k)},0\}).
\end{align}

This problem can actually be avoided by applying the BFGS procedure in the $\NU$-space. The BFGS update in $\NU$-space is obtained by simply putting the $\NU$-space quantities into~\eqref{eq:BFGS}, which yields
\begin{subequations}\label{eq:BFGS_aux_NU}
\begin{align}
B^{(k+1)}_{\NU} &= B^{(k)}_{\NU} - 
\frac{B^{(k)}_{\NU} s_{\NU}^{(k)}s_{\NU}^{(k)T} B^{(k)}_{\NU}}{s_{\NU}^{(k)T} B^{(k)}_{\NU} s_{\NU}^{(k)}} + 
\frac{y_{\NU}^{(k)}y_{\NU}^{(k)T}}{s_{\NU}^{(k)T} y_{\NU}^{(k)}} , \\
\intertext{which is then transformed to $\x$-space via the chain rule}
B^{(k+1)}_{\x} &= \tilde{B}^{(k)}_{\x} - 
\frac{\bar{B}^{(k)}_{\NU} \bar{s}^{(k)}_{\x}\bar{s}^{(k)T}_{\x} \bar{B}^{(k)}_{\NU}}{\bar{s}^{(k)T}_{\x} \bar{B}^{(k)}_{\NU} \bar{s}^{(k)}_{\x}} + 
\frac{\bar{y}^{(k)}_{\x}\bar{y}^{(k)T}_{\x}}{\bar{s}^{(k)T}_{\x} \bar{y}^{(k)}_{\x}} ,
\end{align}
\end{subequations}
where \(\bar{s}^{(k)}_{\x} \isDefinedAs J^{(k+1)-1}s_{\NU}^{(k)}\) and \(\bar{y}^{(k)}_{\x} \isDefinedAs J^{(k+1)}y_{\NU}^{(k)}\).

\begin{figure*}[t!]
        \centering
        \includegraphics[width=\figsize]{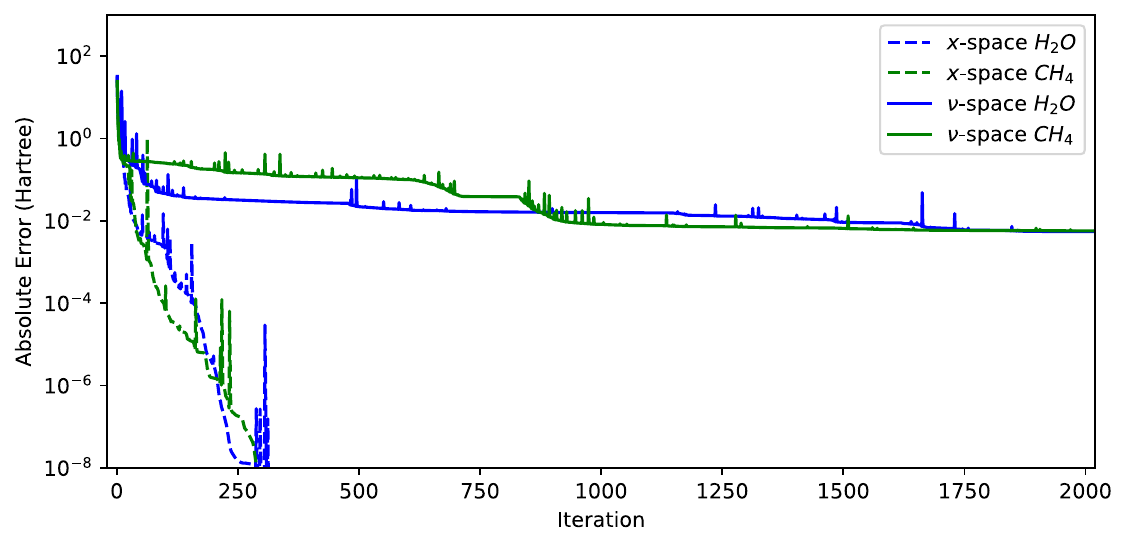}
	\caption{Convergence of the energy ($E^{(k)}-E^{(\text{ref})}$ with respect to \microiteration count) using~\eqref{eq:BFGS_aux_NU} in solid lines and~\eqref{eq:BFGSaux_closeX_secantX} in dashed lines for the \ce{H2O} and \ce{CH4} molecules. }
	\label{fig:conv_H_aux_nu}
\end{figure*}

Though this update could be implemented without additional approximations, the BFGS update in the $\NU$-space actually leads to more convergence problems than~\eqref{eq:BFGSaux_closeX_secantX} as we demonstrate in \Fig~\ref{fig:conv_H_aux_nu}. We suspect that the issue here is that the secant equation in $\NU$-space is less relevant than the secant equation in $\x$-space, as we aim to make a quadratic model in $\x$-space.

\section{Termination criteria for the 2-step procedure}
\label{sec:2step-termination}

\begin{figure*}[t!]
        \centering
        \includegraphics[width=\figsize]{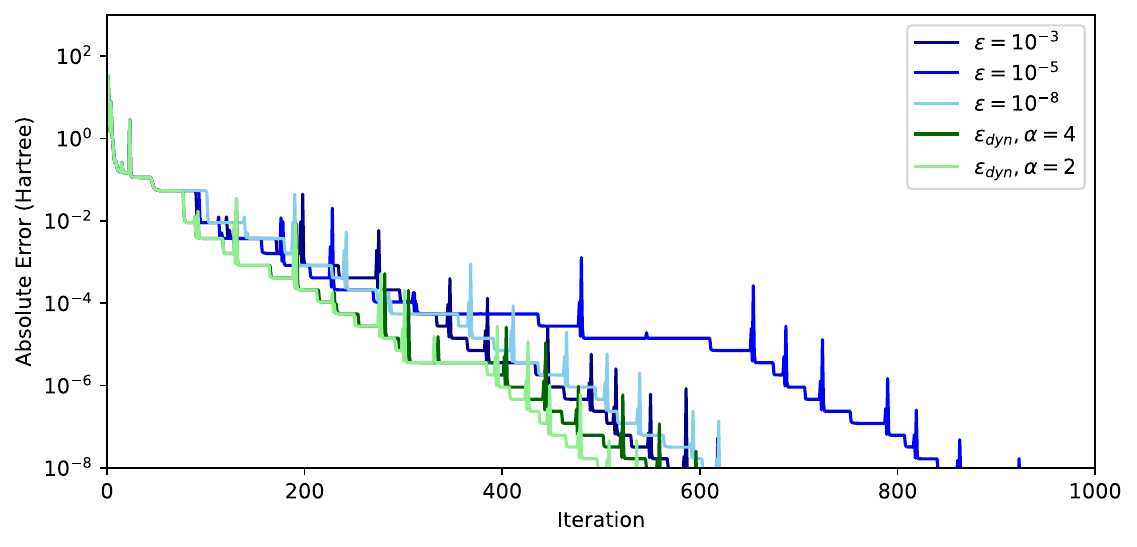}
	\caption{Convergence of the energy ($E^{(k)}-E^{(\text{ref})}$ with respect to the sum of ON and NO \microiteration counts) using different $\epsilon$ termination criteria for the optimi\sation of the NO at each \macroiteration, for the \ce{H2O} molecule, and exact \hessian. For $\epsilon_{\text{dyn}}$, we start at $\epsilon_{\text{dyn}}^{(0)}=10^{-2}$ and take $\epsilon_{\text{dyn}}^{(k+1)} =\frac{1}{\alpha} \epsilon^{(k)}_{\text{dyn}}$ after each \macroiteration until $\epsilon^{(k)}_{\text{dyn}}$ reaches the same precision as the global convergence criterion. }
	\label{fig:epsi_conv_exa}
\end{figure*}

\begin{figure*}[t!]
        \centering
        \includegraphics[width=\figsize]{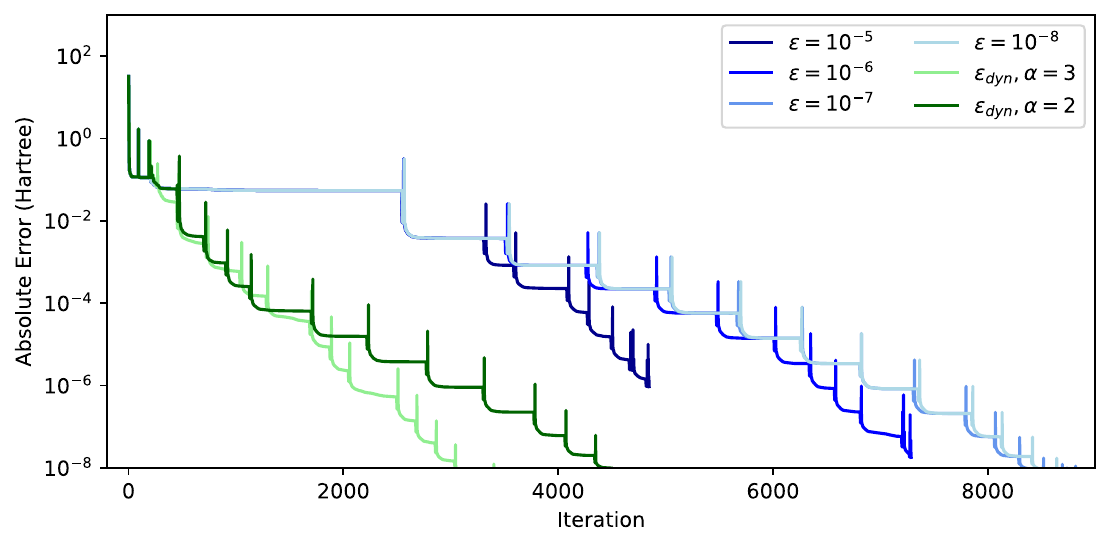}
	\caption{Convergence of the energy ($E^{(k)}-E^{(\text{ref})}$ with respect to the sum of ON and NO \microiteration counts) using different $\epsilon$ termination criteria for the optimi\sation of the NO at each \macroiteration, for the \ce{H2O} molecule, and BFGS as \hessian approximation. For $\epsilon_{\text{dyn}}$, we start at $\epsilon_{\text{dyn}}^{(0)}=10^{-2}$ and take $\epsilon_{\text{dyn}}^{(k+1)} =\frac{1}{\alpha} \epsilon^{(k)}_{\text{dyn}}$ after each \macroiteration until $\epsilon^{(k)}_{\text{dyn}}$ reaches the same precision as the global convergence criterion. }
	\label{fig:epsi_conv_BFGS}
\end{figure*}

In the case of a 2-step procedure, to avoid the waste of NO iterations, a straightforward way would be to adapt the termination condition on the NO optimi\sation while keeping a strong termination criterion on the global convergence. A criterion on the gradient seems more reasonable, yet to emphasis our point, in this section we take a termination condition on the step length $\norm{s_{NO}}<\epsilon$,(where $\epsilon$ is the required termination criterion for one optimi\sation of the NOs), which will exaggerate the problem. We present the results in \Fig~\ref{fig:epsi_conv_exa} and~\ref{fig:epsi_conv_BFGS} with the exact \hessian and BFGS approximation respectively.

When used with a poor \hessian approximation (BFGS here) this approach leads to convergence problems (the darker blue lines in \Fig~\ref{fig:epsi_conv_BFGS} did not reach the \macroiteration termination criterion), when the termination criterion on the NOs, $\epsilon$ is to loos, the algorithm fails to converge. On the other hand, a too strong criterion (i.e. small $\epsilon$) forces the algorithm to compute many ineffective iterations (thousands of those in the lighter blue lines in \Fig~\ref{fig:epsi_conv_BFGS}).

This problem is readily remedied by a dynamic termination condition for the termination threshold $\epsilon_{\text{dyn}}$ of the NO optimi\sation. The dynamic NO termination criterion $\epsilon_{\text{dyn}}$ would start at a large value and be reduced (for example, after each \macroiteration) to a sufficiently small value to ensure global convergence. As simple demonstration we have chosen the update \(\epsilon^{(k+1)} = \epsilon^{(k)}/\alpha\). In \Fig~\ref{fig:epsi_conv_exa} and~\ref{fig:epsi_conv_BFGS} we show the energy convergence for \(\epsilon^{(0)} = 10^{-2}\) and both $\alpha = 2$ and $\alpha = 3$ in green. All values for $\alpha$ (green lines) are more efficient than the static NO convergence criteria (blue lines), especially for the BFGS case (\Fig~\ref{fig:epsi_conv_BFGS}), but remain inefficient compared to a 1-step procedure (see dark blue line in \Fig~\ref{fig:exa_hess_conv} and \ref{fig:BFGS_conv}). 

\begin{figure*}[t!]
        \centering
        \includegraphics[width=\figsize]{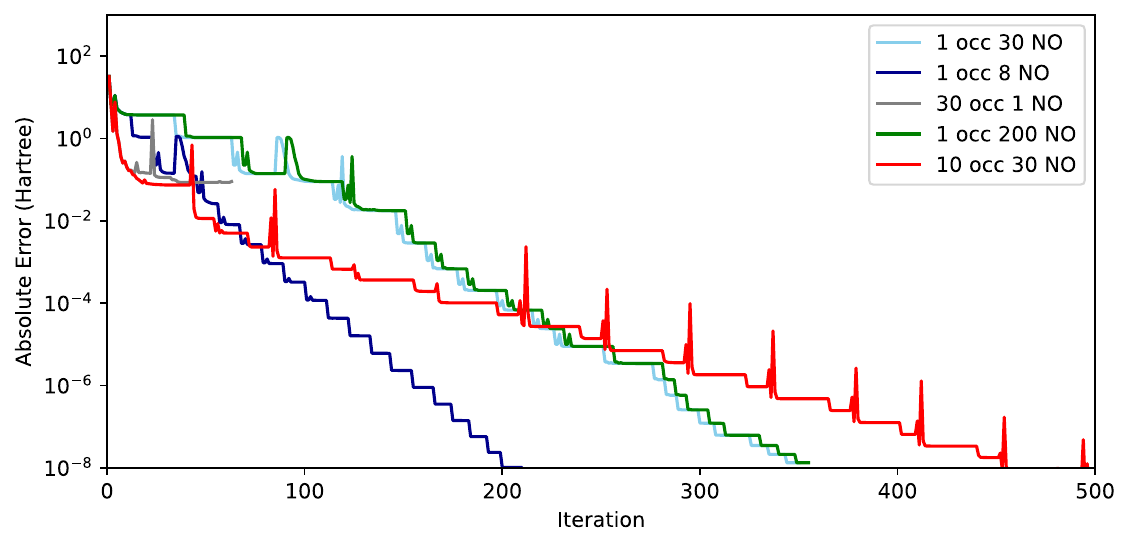}
	\caption{Convergence of the energy ($E^{(k)}-E^{(\text{ref})}$ with respect to the sum of ON and NO \microiteration counts) using different limits on the number of iteration for the optimi\sation of the NO and ON at each \macroiteration, for the \ce{H2O} molecule, and exact \hessian. The first number indicates the maximum number of ON iteration per \macroiteration and the second the maximum number of NO per \macroiteration. }
	\label{fig:iter_conv_exa}
\end{figure*}

Another criterion that can be considered, is the number of iteration allowed per \macroiteration.\cite{Lew-YeeFelipe2023} The convergence is reported for some set of maximum ON and NO iterations in \Fig~\ref{fig:iter_conv_exa}. We first notice that the number NO iteration cannot be to small, otherwise the update do not sufficiently change the energy and the algorithm stops (grey line). We can also notice that changing the maximum number of iteration can significantly impact the final number of iterations (purple line needs around 200 iterations to converge to $10^{-8}$ Hartree, while the red line, more than 400), and that a small number of NO iterations per ON iteration is more efficient. This indeed corresponds to limiting the amount of inefficient NO iterations. 

\section{Other separable functionals}
\subsection{The Power functional with $\alpha=0.7$}

\begin{figure*}[t!]
    \centering
    \settototalheight{\dimen0}{\includegraphics[width=\figsize]{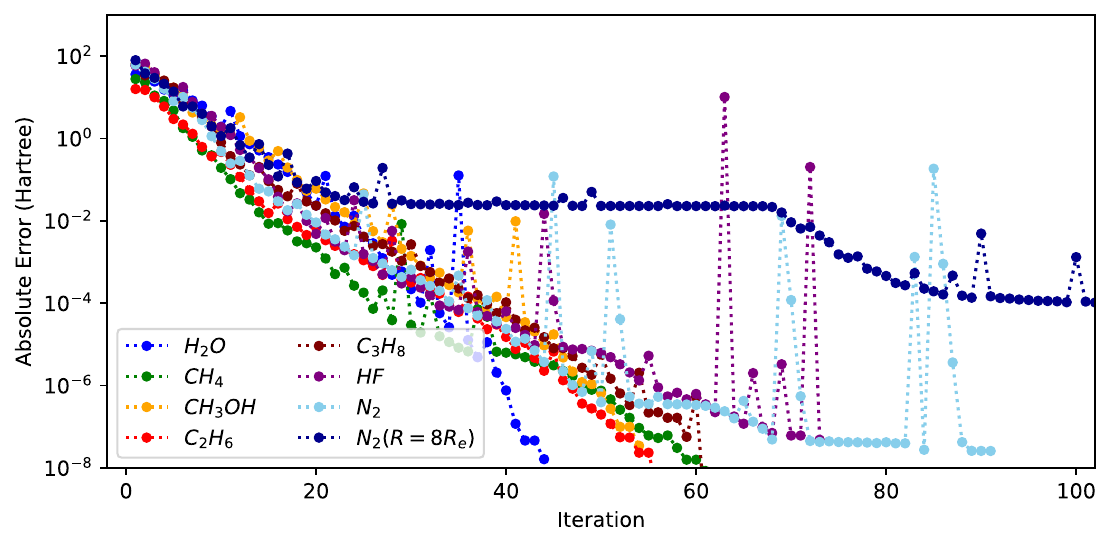}}%
    \includegraphics[width=\figsize]{exa_hess_conv_Power_svg-tex.pdf}%
    \llap{\raisebox{\dimen0-3cm}{
      \includegraphics[height=3cm]{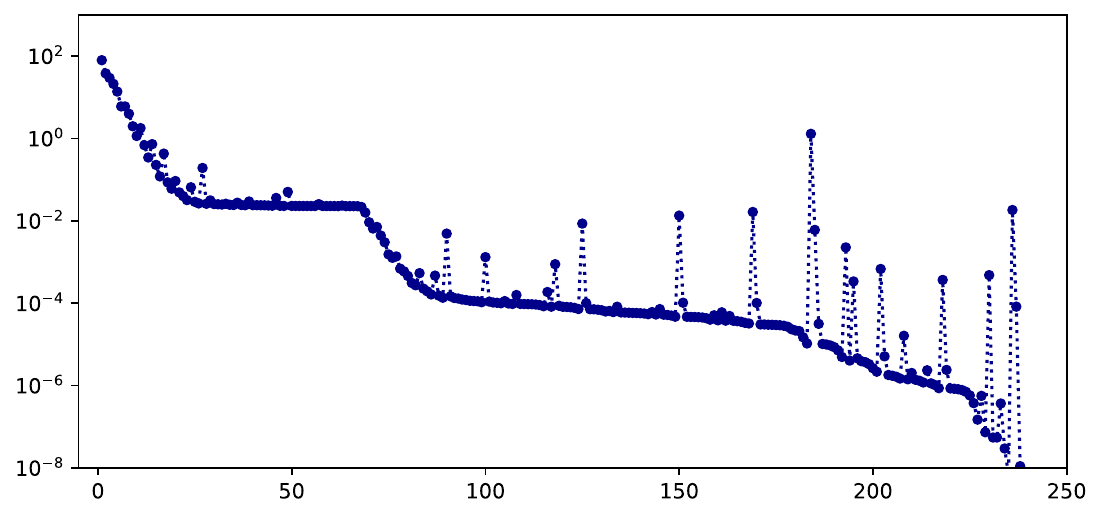}%
    }}
	\caption{Convergence of the energy ($E^{(k)}-E^{(\text{ref})}$ given by the Power functional ($\alpha=0.7$) with respect to \microiteration count) using the exact \hessian.}
	\label{fig:exa_hess_power}
\end{figure*}

\begin{figure*}[t!]
    \centering
        \includegraphics[width=\figsize]{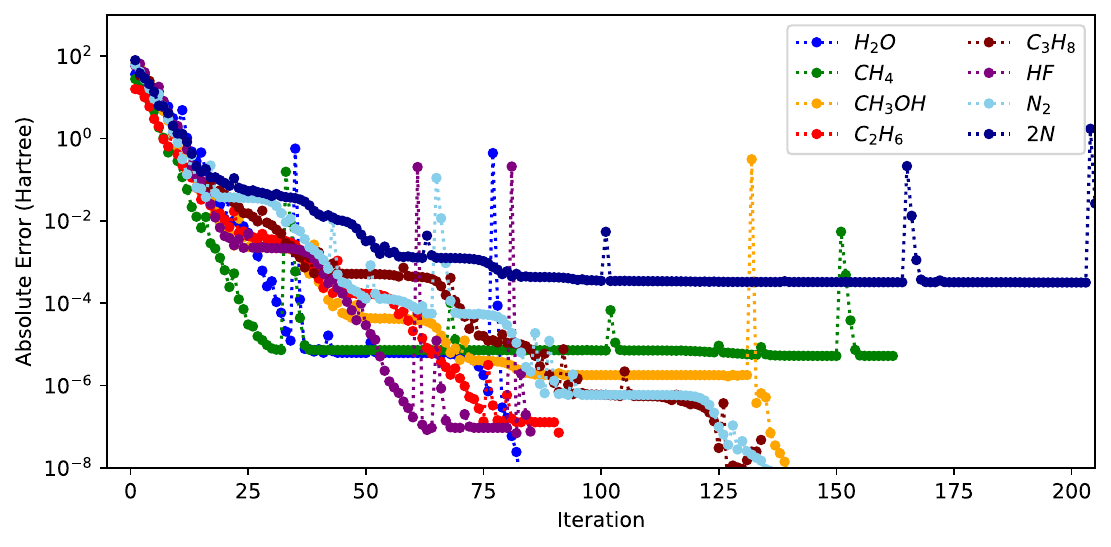}
	\caption{Convergence of the energy ($E^{(k)}-E^{(\text{ref})}$ given by the Power functional ($\alpha=0.7$) with respect to \microiteration count) using $H^{\cheap}_{\x}$ exactly and equation~\eqref{eq:xBFGS_aux_x} to approximate $H^{\exp}$ with the prefactor $A(s_n,10^{-3})$ (see text in the article).}
	\label{fig:conv_tBFGS_Power}
\end{figure*}

\begin{figure*}[t!]
        \centering
        \includegraphics[width=\figsize]{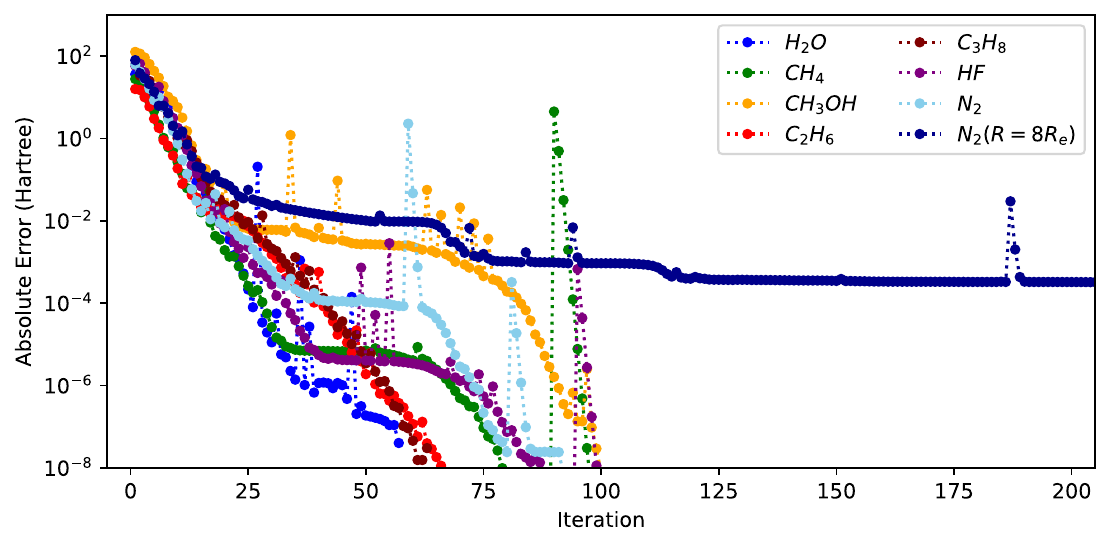}
	\caption{Convergence of the energy ($E^{(k)}-E^{(\text{ref})}$ given by the Power functional ($\alpha=0.7$) with respect to \microiteration count) using $H^{\cheap}_{\x}$ exactly and equation~\eqref{eq:nuBFGS_aux_x} to approximate $H^{\exp}$ with the prefactor $A(s_n,10^{-3})$ (see text in the article). }
	\label{fig:conv_bBFGS_Power}
\end{figure*}

Our approach requires additional work to be adapted to non-separable functionals, but can readily be employed for separable functionals other than Müller. Unfortunately, there are few of then, all taking the form, 
\begin{align}
    W^{\text{Power}}[\gamma] &= \frac{1}{2}\sum_{ij} n_i n_j [ii|jj] \notag\\*
    &- \frac{1}{2}\sum_{ij} (n_i n_j)^{\alpha} [ij|ji],
\end{align}
where $\alpha\in (0,1]$. The Müller functional corresponds to the case $\alpha=1/2$ (studied in the article), the Hatree--Fock functional (see \Section~\ref{subsec:HF_func}), $\alpha=1$, but other values of $\alpha$ have been proposed in the literature.\cite{SharmaDewhurst2008, LathiotakisSharma2009, PutajaRasanen2011, KamilShade2016} As an example, here we take $\alpha=0.7$ and test our algorithm with the exact \hessian in \Fig~\ref{fig:exa_hess_power}, approximations~\eqref{eq:nuBFGS_aux_x} and~\eqref{eq:xBFGS_aux_x} (with $H^{\cheap}$ exactly and damping factor as in \Section~\ref{sec:approx_hess} of the article) in \Fig~\ref{fig:conv_tBFGS_Power} and \Fig~\ref{fig:conv_bBFGS_Power} respectively.

With the exact \hessian, the number of iterations to reach a precision of $10^{-8}$ Hartree is similar to the case of the Müller functional (see \Fig~\ref{fig:exa_hess_conv} of the article) for most molecules, but the convergences display more unsuccessful steps (energy jumps) for diatomic molecules and \ce{C3H8}, and fails to converge beyond an error of $10^{-7}$ (purple and light blue lines of \Fig~\ref{fig:exa_hess_power}). The algorithm moreover struggles to converge for the strong correlated system (dark blue line), and only reaches the convergence criterion after 240 iterations.

The approximation~\eqref{eq:xBFGS_aux_x} seems to work worse with the Power functional, it fails to converge for a certain number of molecules, including \ce{CH4} which did not prove to be challenging up to this point (see \Fig~\ref{fig:conv_tBFGS_Power}). Fortunately, the update~\eqref{eq:nuBFGS_aux_x} gives convergences comparable to the case of the Müller functional (\Fig~\ref{fig:conv_bBFGS_Power} compared to \Fig~\ref{fig:mixed_thess_conv}), and most molecule reach an error of $10^{-7}$ to $10^{-8}$ Hartree (as the exact \hessian, \Fig~\ref{fig:exa_hess_power}) within 200 iterations. A notable exception is the stretched \ce{N2} molecule, which did not converge, but this system was already challenging for the exact \hessian. 

\subsection{The Hartree--Fock functional}
\label{subsec:HF_func}
\begin{figure*}[t!]
        \centering
        \includegraphics[width=\figsize]{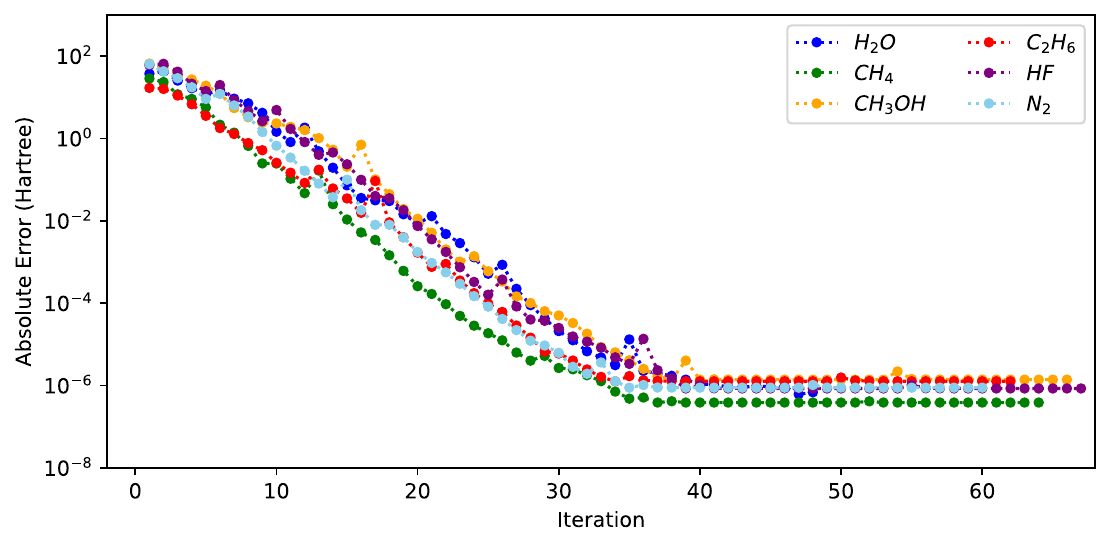}
	\caption{Convergence of the energy ($E^{(k)}-E^{(\text{ref})}$ given by the Hartree--Fock functional with respect to \microiteration count) using the exact \hessian.}
	\label{fig:conv_exa_HF}
\end{figure*}

\begin{figure*}[t!]
        \centering
        \includegraphics[width=\figsize]{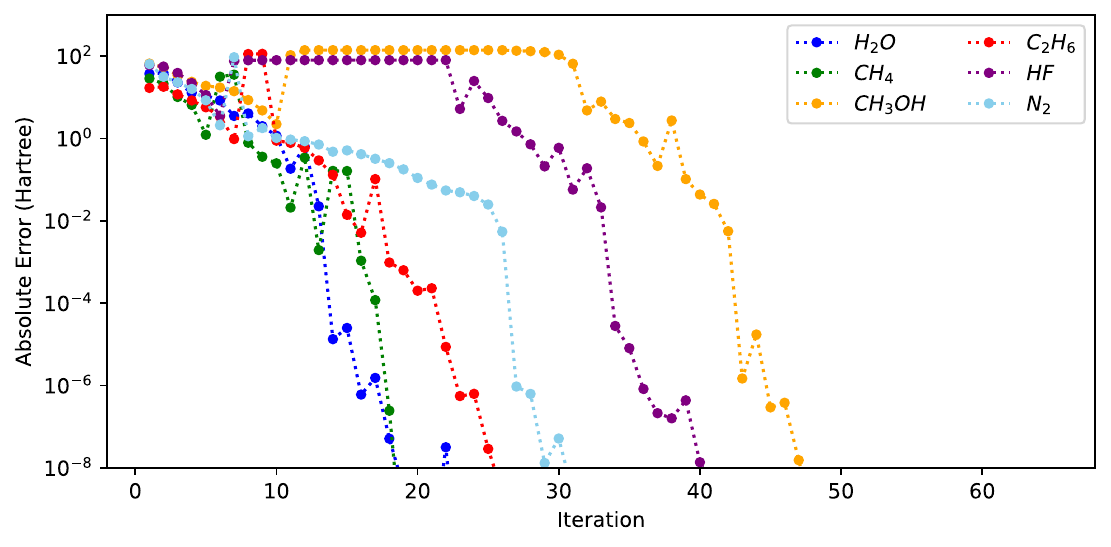}
	\caption{Convergence of the energy ($E^{(k)}-E^{(\text{ref})}$ given by the Hartree--Fock functional with respect to \microiteration count) using the exact \hessian. Here the boundary constraints on the ONs were handle using~\eqref{eq:CancesPernal_aparam} instead of EBI. Note that the energy jumps for \ce{HF} and \ce{C3H8} around 10 iterations are due to an instability in the code, which sometimes fails to impose the trace constraint $\Tr\{\gamma\}=N$.}
	\label{fig:conv_exa_HF_CP}
\end{figure*}

\begin{figure*}[t!]
        \centering
        \includegraphics[width=\figsize]{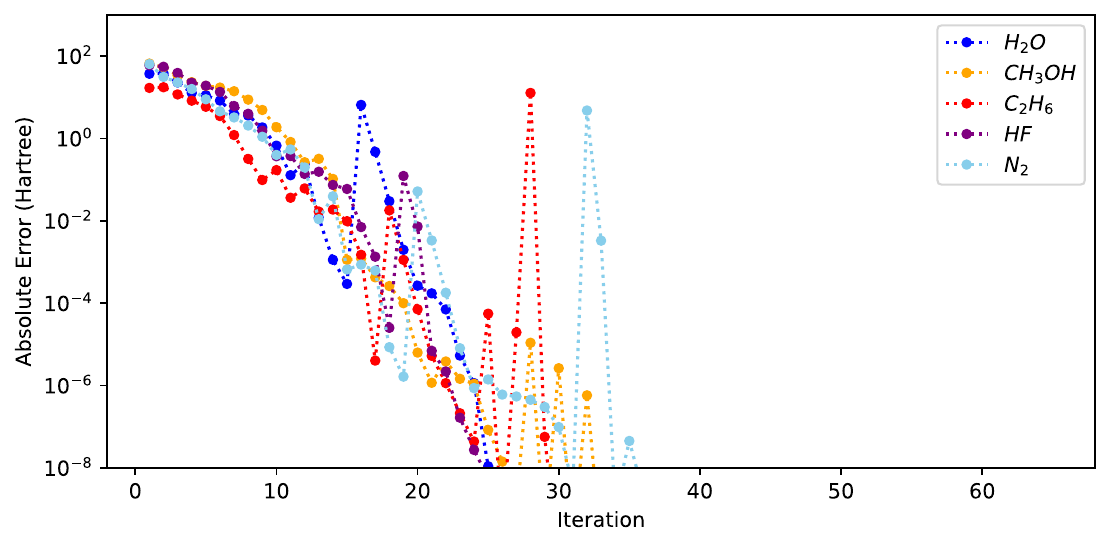}
	\caption{Convergence of the energy ($E^{(k)}-E^{(\text{ref})}$ given by the Hartree--Fock functional with respect to \microiteration count) using $H^{\cheap}_{\x}$ exactly and~\eqref{eq:nuBFGS_aux_x} to approximate $H^{\exp}$ with the prefactor $A(s_n,10^{-3})$ (see text in the article). Here the boundary constraints on the ONs were handle using~\eqref{eq:CancesPernal_aparam} instead of EBI. The result for \ce{CH4} was not reported has is failed to impose the trace constraint $\Tr\{\gamma\}=N$.}
	\label{fig:conv_bBFGS_HF_CP}
\end{figure*}

Although, not practically relevant it is possible to define a functional in 1RDMFT, corresponding (i.e. giving the same ground state 1RDM and energy) to the Hartree--Fock method. On the contrary to the Müller functional, the Hartree--Fock is concave in the space of 1RDM,\cite{Lieb1981,VeeraraghavanMazziotti2014} which makes it a relevant test for our method. We thus report the energy convergence with the Hartree--Fock functional when using the exact \hessian in \Fig~\ref{fig:conv_exa_HF}. While the algorithm converges quiet well to an error of $10^{-6}$ Hartree for all the tested molecules, it fails to improve beyond that precision. The reason is that the error function in the EBI parametri\sation tends to be very flat when the ONs tend to 0 or 2, which makes the gradient and \hessian vanish, preventing the algorithm from converging further. A way around it is to use an other parametri\sation associated with a projection of the ONs as proposed by Canc\`es and Pernal,\cite{CancesPernal2008}
\begin{equation}\label{eq:CancesPernal_aparam}
    \sqrt{n}_i= 
    \begin{cases}
        \sqrt{2}\frac{x_i}{1+\nu} & \text{ if } x_i \in [0,1+\nu] \\
        0 & \text{ if } x_i<0 \\
        \sqrt{2} & \text{ if } x_i>1+\nu,
    \end{cases}
\end{equation}
where $x_i\in \mathds{R}$ and $\nu\in\mathds{R}^+\text{ s.t. }\Tr\{\gamma\}\leq N_e$. 
We have tested our algorithm while replacing only the way we handle constraints on the ONs i.e.\ EBI by~\eqref{eq:CancesPernal_aparam} and report the results in \Fig~\ref{fig:conv_exa_HF_CP}.
Aside from some instabilities due to the immaturity of the code, the approach of equation~\eqref{eq:CancesPernal_aparam} seems to indeed solve the problem and allow the algorithm to converge decently fast to $10^{-8}$ Hartree. It has also been possible to test the approximation of \Section~\ref{sec:approx_hess} of the article with the Hartree--Fock functional and using~\eqref{eq:CancesPernal_aparam}, which give a convergence comparable to the one of the exact \hessian (up to different jumps due to the instabilities), as shown in \Fig~\ref{fig:conv_bBFGS_HF_CP}.

\section{Derivation of the BFGS approximation}
\label{sec:appendix_BFGS}

The idea of (quasi-)Newton methods is to take the derivative of a second order model $m$ of the function $f$ to optimise, which writes for the $k^{th}$ iteration
\begin{equation}
    \grad m^{(k)}(p) = \grad f^{(k)} + B^{(k)} p, 
\end{equation}
where $B$ is the (approximate) \hessian. We then impose that the derivative of $m$ coincides with the derivative of $f$ at the last two iterations. For the last iterations it is automatically verified by taking $p=0$, and for the second last iterations we have to take $p=-s^{(k-1)}$, that is, the opposite of the last step, and we get
\begin{equation}
    \grad m^{(k)}(-s^{(k-1)}) = \grad f^{(k)} - B^{(k)} s^{(k-1)} = \grad f^{(k-1)}.
\end{equation}
By defining $y^{(k-1)} = \grad f^{(k)}-\grad f^{(k-1)}$, we obtain the, so called, secant equation,
\begin{equation}\label{eq:sec_eq}
    B^{(k+1)}s^{(k)} = y^{(k)}.
\end{equation}
To obtain the BFGS update~\eqref{eq:BFGS}, we need to solve the problem~\eqref{eq:BFGS_prob}
\begin{multline}\label{eq:BFGS_prob2}
    \min_{B\in \mathds{R}^{N\times N}} \norm{ B-B^{(k)}} \\ \quad\text{ s.t. } \quad B^T = B \text{ and } B s^{(k)} = y^{(k)},
\end{multline}
with the Frobenius norm for $\norm{\cdot}$.

We follow the proof given by Hauser.\cite{HauserOxfordNotes2005} To ensure that $B$ is
symmetric, we write it as $B= LL^T$, $L\in \mathds{R}^{N\times N}$. Taking $g = L^T s^{(k)}$ the secant equation becomes 
\begin{equation}\label{eq:sec_eq_L}
    L g = y^{(k)}.
\end{equation}
Taking the scalar product $\braket{A}{B} = \Tr(A B^T)$, \eqref{eq:BFGS_prob2} is equivalent to
\begin{multline}
    \min_L \braket{L-L^{(k)}}{L-L^{(k)}} \quad \\ \text{s.t.} \quad \braket{L}{e_i g^T} = y^{(k)}_i \forall i,
\end{multline}
with $e_i$ the $i^{th}$ column of the identity matrix, that is $e_i g^T$ are the normal vectors defining the affine subspace $\mathcal{P}_g$ in which $L$ has to be optimi\sed. The minimizer $L^*$ is the closest point to $L^{(k)}$ in $\mathcal{P}_g$. Thus, $L^*-L^{(k)}$ is orthogonal to $\mathcal{P}_g$ and so a linear combination of the $e_i g^T$.~\eqref{eq:BFGS_prob2} is then equivalent to find $L^* \in \mathds{R}^{N\times N}$ and $\lambda^* \in \mathds{R}^N$ satisfying
\begin{subequations}
\begin{align}
    \label{eq:LdifEq}
    (L^*-L^{(k)}) - \lambda^*g^T &=0 \\
    \label{eq:Lplane}
    L^* g &= y^{(k)}.
\end{align}
\end{subequations}
Multiplying~\eqref{eq:LdifEq} from the right by $g$ and utilizing~\eqref{eq:Lplane} gives
\begin{subequations}\label{eq:BFGS_sol}
\begin{align}
    \lambda^* &= \frac{y^{(k)}-L^{(k)}g}{g^Tg} , \\
\intertext{which can be inserted back to yield}
    L^* &= L^{(k)} + \frac{(y^{(k)}-L^{(k)}g)g^T}{g^Tg} .
\end{align}
\end{subequations}
Inserting the minimizer $L^*$ in the definition of $g$ and rearranging gives
\begin{equation}
    L^{(k)T} s^{(k)} = g\left(1-\frac{(y^{(k)}-L^{(k)}g)^T s^{(k)}}{g^Tg}\right),
\end{equation}
meaning that $g \propto L^{(k)T}s^{(k)}$. We take $g = \beta L^{(k)T} s^{(k)}$ and by inserting it into the previous equation, we get $\beta = \pm \sqrt{\frac{y^{(k)T}s^{(k)}}{s^{(k)T} B^{(k)} s^{(k)}}}$. We now have an explicit expression for $L^*$ and simplifying $B^{(k+1)} \equiv B^* \equiv L^*L^{*T}$ we get the BFGS approximation~\eqref{eq:BFGS}.

\bibliography{bibliography.bib}